\documentclass[preprint,showpacs,preprintnumbers,amsmath,amssymb,superscriptaddress,12pt,graphicx]{revtex4}

\pdfoutput=1

\usepackage{graphicx}
\usepackage{dcolumn}
\usepackage{bm}
\usepackage{amssymb}
\usepackage{color}
\usepackage{multirow}

\newcommand{\tc}{\textcolor{black}}

 \def\tskip{\setlength{\tskip}{5pt}}
\def\colwidth{\setlength{\colwidth}{3.5in}}

\def\prd{Phys. Rev. D}

\def\prl{Phys. Rev. Lett.~}
\def\plb{Phys. Lett. B}
\def\jcap{JCAP~}
\def\apj{Astrophys. J.~}

\def\mnras{Mon. Not. Roy. Astron. Soc.~}
\def\apjs{Astrophys. J. Suppl. Ser.~}
\def\aanda{Astron. Astrophys. ~}

\def\ijmpd{Int. J. Mod. Phys. D}

\newcommand{\lsim}{\mathrel{\hbox{\rlap{\lower.55ex\hbox{$\sim$}} \kern-.3em \raise.4ex \hbox{$<$}}}}
\newcommand{\gsim}{\mathrel{\hbox{\rlap{\lower.55ex\hbox{$\sim$}} \kern-.3em \raise.4ex \hbox{$>$}}}}
\newcommand{\beq}{\begin{equation}}
\newcommand{\eeq}{\end{equation}}
\newcommand{\be}{\begin{equation}}
\newcommand{\ee}{\end{equation}}
\newcommand{\bes}{\begin{equation*}}
\newcommand{\ees}{\end{equation*}}
\newcommand{\beqa}{\begin{eqnarray}}
\newcommand{\eeqa}{\end{eqnarray}}
\newcommand{\bea}{\begin{eqnarray}}
\newcommand{\ena}{\end{eqnarray}}

\begin{document}

\title{Directional dependence of CMB parity asymmetry}

\author{Wen Zhao\footnote{Email: wzhao7@ustc.edu.cn}}
\affiliation{Department of Astronomy, University of Science and
Technology of China, Hefei, 230026, China \\ Key Laboratory for
Researches in Galaxies and Cosmology, University of Science and
Technology of China, Hefei, 230026, China }

\date{\today}

\begin{abstract}

{Parity violation in the cosmic microwave background (CMB)
radiation has been confirmed by recent Planck observation. In
this paper, we extend our previous work [P. Naselsky et al.
Astrophys. J. {\bf 749}, 31 (2012)] on the directional properties
of CMB parity asymmetry by considering the Planck data of the CMB
temperature anisotropy. We define six kinds of the directional
statistics and find that they all indicate the odd-parity
preference of CMB data. In addition, we find the preferred axes of
all these statistics strongly correlate with the preferred axes of
the CMB kinematic dipole, quadrupole, and octopole. The alignment between them is confirmed
at more than $3\sigma$ confidence level,
which implies that the CMB parity asymmetry, and the anomalies of
the CMB quadrupole and octopole, may have the common physical,
contaminated or systematic origin. In addition, they all should be connected with the possible contamination of the residual dipole component.}

\end{abstract}



\maketitle


\section{Introduction \label{section1}}

The modern cosmological model is based on the cosmological
principle: the Universe is homogeneous and isotropic on Hubble
scales, which has been confirmed by various observations mainly
from the isotropy of the cosmic microwave background (CMB)
radiation \cite{wmap,planck}. The recently released Planck
data on the CMB temperature anisotropy are excellently consistent with
the base $\Lambda$CDM model, especially at the high multipoles
$l>40$ \cite{planck2}. However, similar to the WMAP
data \cite{wmap2}, a number of anomalies has been reported in the CMB
low multipoles \cite{quadrupole}, including the low quadrupole,
the alignment of the quadrupole and octopole, the missing angular power at
large scale and so on. Among them, the parity asymmetry of the CMB
low multipoles has also been investigated in both WMAP and
Planck data \cite{land,kim,
kim2,gruppuso,hansen,aluri,quadrupole}, showing significant
dominance of the power spectrum stored in the odd multipoles over
the even ones. This anomaly of CMB may imply the physics of the
early Universe \cite{inflation,inflation2}, the nontrivial
topology \cite{topology,topology2}, some foreground residuals
\cite{hansen,burigana} or systematic errors \cite{system}.

The connections between these anomalies have been investigated
by several groups, which are helpful to reveal the physics behind them.
For example, it was shown that the low
quadrupole and the odd-multipole preferences were tidily connected with
the lack of the two-point correlation functions in both scales
$60^{\circ}\le \Theta \le 180^{\circ}$ \cite{schwarz, copi} and
$1^{\circ} \le \Theta \le 30^{\circ}$ \cite{kim2011}. In
Ref. \cite{naselskyzhao}, we have studied the directional properties of the CMB
parity asymmetry in the WMAP data and found the preferred
directions always coincide with the CMB kinematic dipole.
Recently, the Planck team has released the CMB products SMICA
{\tc{(an implementation of independent component analysis
of power spectra)}}, NILC {\tc{(a needlet-based version of internal
linear combination)}}, and SEVEM {\tc{(template fitting using
the lowest and highest freuqency bands)}} maps,
which are made via quite different techniques \cite{planckdata}.
In this paper, applying Planck data to all these, we shall extend
this work by defining six kinds of directional parity statistics.
We find the preferred axes of all these statistics are quite close
to each other. This indicates that the preferred axis of the
CMB parity asymmetry really exists, which is independent of
definitions of the statistics.  Comparing with the preferred axes
of the CMB quadrupole and octopole, as well as the kinematic
dipole, we find the preferred axis of CMB parity
asymmetry is very close to all of them. This
study suggests that the CMB parity violation may have the same
origin with the anomalies of CMB quadrupole and octopole, which might connect with the CMB kinematic dipole.

The outline of the paper is as follows. In Sec. 2, we
introduce the basic characteristics of the CMB parity asymmetry
and study the orientations of maximum parity asymmetry by defining
various directional statistics. In Sec. 3, we compare this
preferred axis with those of CMB kinematic dipole, quadrupole, and octopole and find the
alignment between them. In Sec. 4, we summarize our
investigations.

\section{CMB parity asymmetry and the directional dependence \label{section2}}

The CMB temperature fluctuations on a sphere are usually decomposed as,
 \be
 \Delta T(\theta,\phi)=\sum_{l=0}^{\infty} \sum_{m=-l}^{l} a_{lm} Y_{lm}(\theta,\phi),
 \ee
where $Y_{lm}(\theta,\phi)$ are the spherical harmonics and
$a_{lm}$ are the corresponding coefficients. Under the assumption
of the random Gaussian field, the amplitudes $|a_{lm}|$ are
distributed according to Rayleigh's probability distribution
function, and the phase of $a_{lm}$ with $m\neq 0$ is supported to
be evenly distributed in the range $[0,2\pi]$. The power spectrum
of CMB is defined as
 \be\label{cl}
 C_l \equiv \frac{1}{2l+1} \sum_{m=-l}^{l} |a_{lm}|^2,
 \ee
which is rotationally invariant; i.e., the power spectrum in Eq.
(\ref{cl}) is invariant for any rotations of the reference system
of the coordinate.

\begin{figure*}[t]
\begin{center}
\includegraphics[width=10cm]{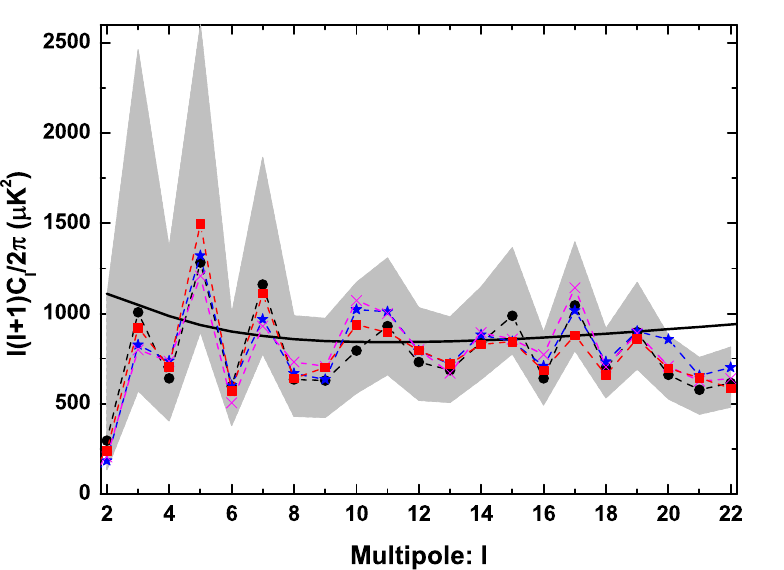}
\end{center}
\caption{{ The CMB power spectra at low multipoles. The black \emph{dots} are obtained from the masked Planck C-R map, and the shadow regions are the corresponding $68\%$ ranges on the posteriors, which are all given in                                                                                                                                                                                                                                                                                                                                                                                                                                                                                                                                                                                                                                                                                                                                                                                                                                                                                                                                                                                                                                                                                                                                                                                                                                                                                                                                                                                                                                                                                                                                                                                                                                                                                                                                                                                                                                                                                                                                                                                                                                                                                                                                                                                                                                                                                                                                                                                                                                                                                                                                                                                                                                                                                                                                                                                                                                                                                                                                                                                                                                                                                                                                                                                                                                                                                                                                                                                                                                                                                                                                                                                                                                                                                                                                                                                                                                                                                                                                                                                                                                                                                                                                                                                                                                                                                                                                                                                                                                                                                                                                                                                                                                                                                                                                                                                                                                                                                                                                                                                                                                                                                                                                                                                                                                                                                                                                                                                                                                                                                                                                                                                                                                                                                                                                                                                                                                                                                                                                                                                                                                                                                                                                                                                                                                                                                                                                                                                                                                                                                                                                                                                                                                                                                                                                                                                                                                                                                                                                                                                                                                                                                                                                                                                                                                                                                                                                                                                                                                                                                                                                                                                                                                                                                                                                                                                                                                                                                                                                                                                                                                                                                                                                                                                                                                                                                                                                                                                                                                                                                                                                                                                                                                                                                                                                                                                                                                                                                                                                                                                                                                                                                                                                                                                                                                                                                                                                                                                                                                                                                                                                                                                                                                                                                                                                                                                                                                                                                                                                                                                                                                                                                                                                                                                                                                                                                                                                                                                                                                                                                                                                                                                                                                                                                                                                                                                                                                                                                                                                                                                                                                                                                                                                                                                                                                                                                                                                                                                                                                                                                                                                                                                                                                                                                                                                                                                                                                                                                                                                                                                                                                                                                                                                                                                                                                                                                                                                                                                                                                                                                                                                                                                                                                                                                                                                                                                                                                                                                                                                                                                                                                                                                                                                                                                                                                                                                                                                                                                                                                                                                                                                                                                                                                                                                                                                                                                                                                                                                                                                                                                                                                                                                                                                                                                                                                                                                                                                                                                                                                                                                                                                                                                                                                                                                                                                                                                                                                                                                                                                                                                                                                                                                                                                                                                                                                                                                                                                                                                                                                                                                                                                                                                                                                                                                                                                                                                                                                                                                                                                                                                                                                                                                                                                                                                                                                                                                                                                                                                                                                                                                                                                                                                                                                                                                                                                                                                                                                                                                                                                                                                                                                                                                                                                                                                                                                                                                                                                                                                                                                                                                                                                                                                                                                                                                                                                                                                                                                                                                                                                                                                                                                                                                                                                                                                                                                                                                                                                                                                                                                                                                                                                                                                                                                                                                                                                                                                                                                                                                                                                                                                                                                                                                                                                                                                                                                                                                                                                                                                                                                                                                                                                                                                                                                                                                                                                                                                                                                                                                                                                                                                                                                                                                                                                                                                                                                                                                                                                                                                                                                                                                                                                                                                                                                                                                                                                                                                                                                                                                                                                                                                                                                                                                                                                                                                                                                                                                                                                                                                                                                                                                                                                                                                                                                                                                                                                                                                                                                                                                                                                                                                                                                                                                                                                                                                                                                                                                                                                                                                                                                                                                                                                                                                                                                                                                                                                                                                                                                                                                                                                                                                                                                                                                                                                                                                                                                                                                                                                                                                                                                                                                                                                                                                                                                                                                                                                                                                                                                                                                                                                                                                                                                                                                                                                                                                                                                                                                                                                                                                                                                                                                                                                                                                                                                                                                                                                                                                                                                                                                                                                                                                                                                                                                                                                                                                                                                                                                                                                                                                                                                                                                                                                                                                                                                                                                                                                                                                                                                                                                                                                                                                                                                                                                                                                                                                                                                                                                                                                                                                                                                                                                                                                                                                                                                                                                                                                                                                                                                                                                                                                                                                                                                                                                                      the left panel of Fig. 39 in Ref. \cite{planck2}.} The red \emph{squares} are the spectra derived from SMICA map, the blue \emph{stars} are those from NILC map, and the magenta \emph{crosses} are those from the SEVEM map. }\label{fig0}
\end{figure*}

\begin{figure*}[t]
\begin{center}
\includegraphics[width=5cm]{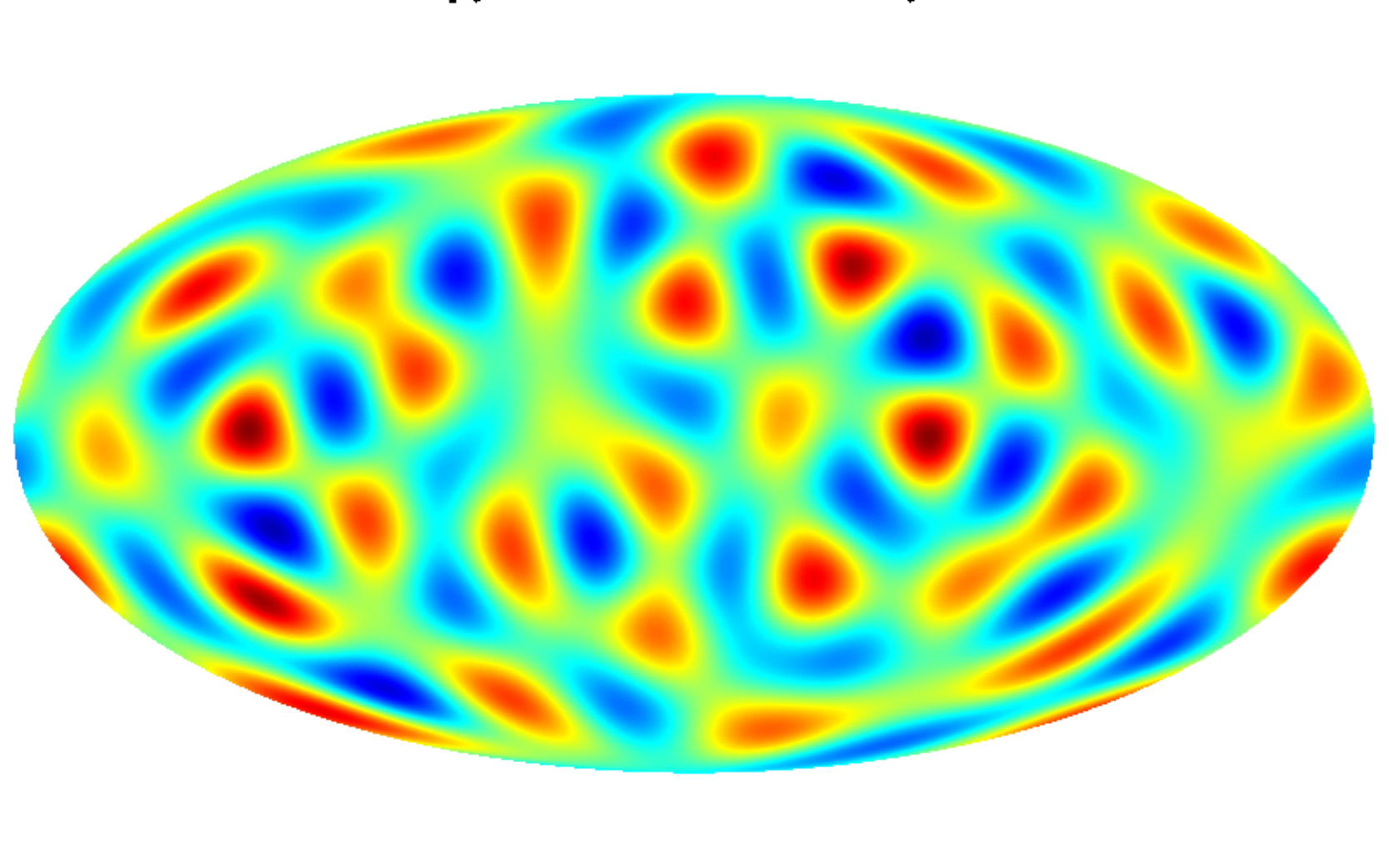}\includegraphics[width=5cm]{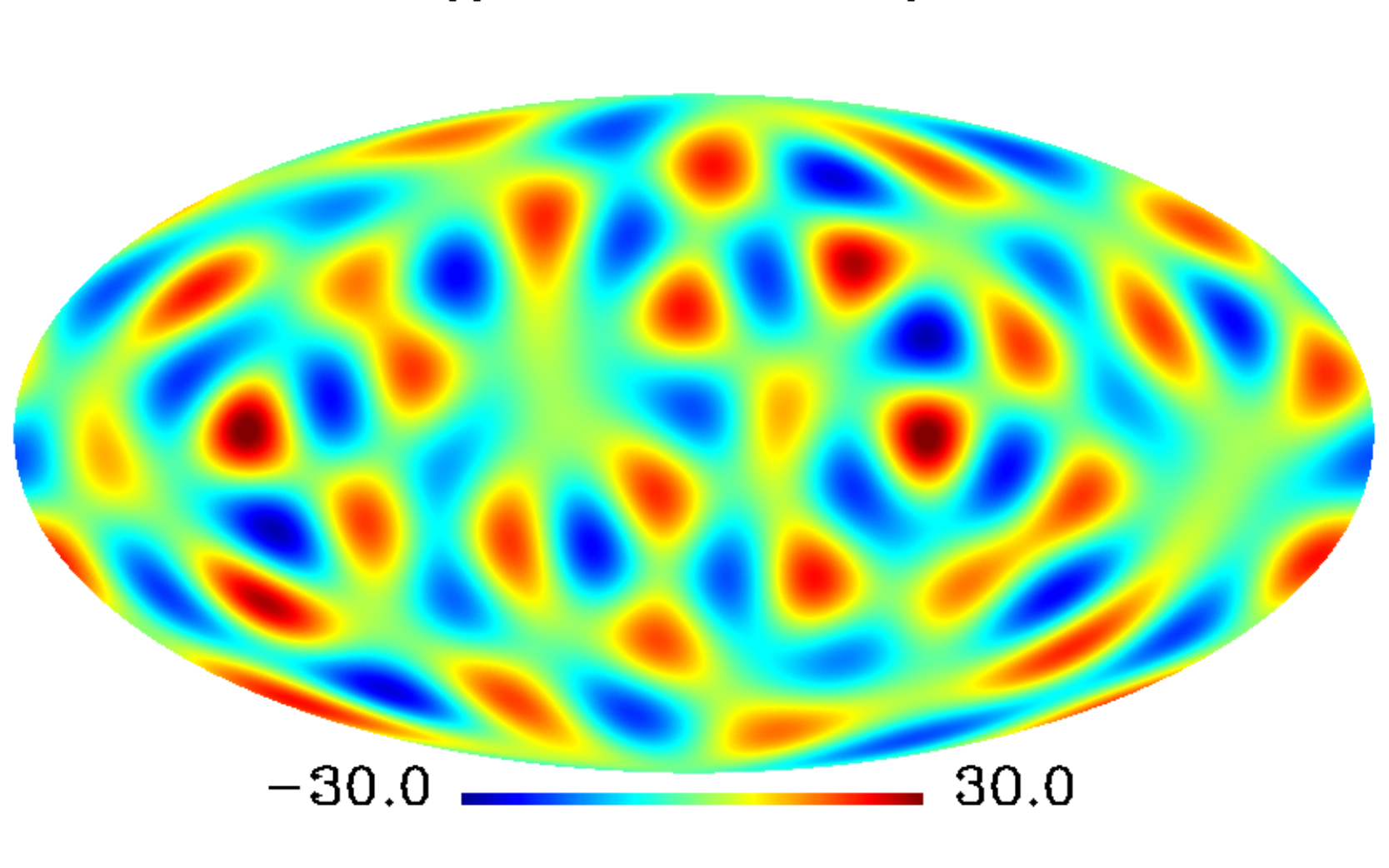}\includegraphics[width=5cm]{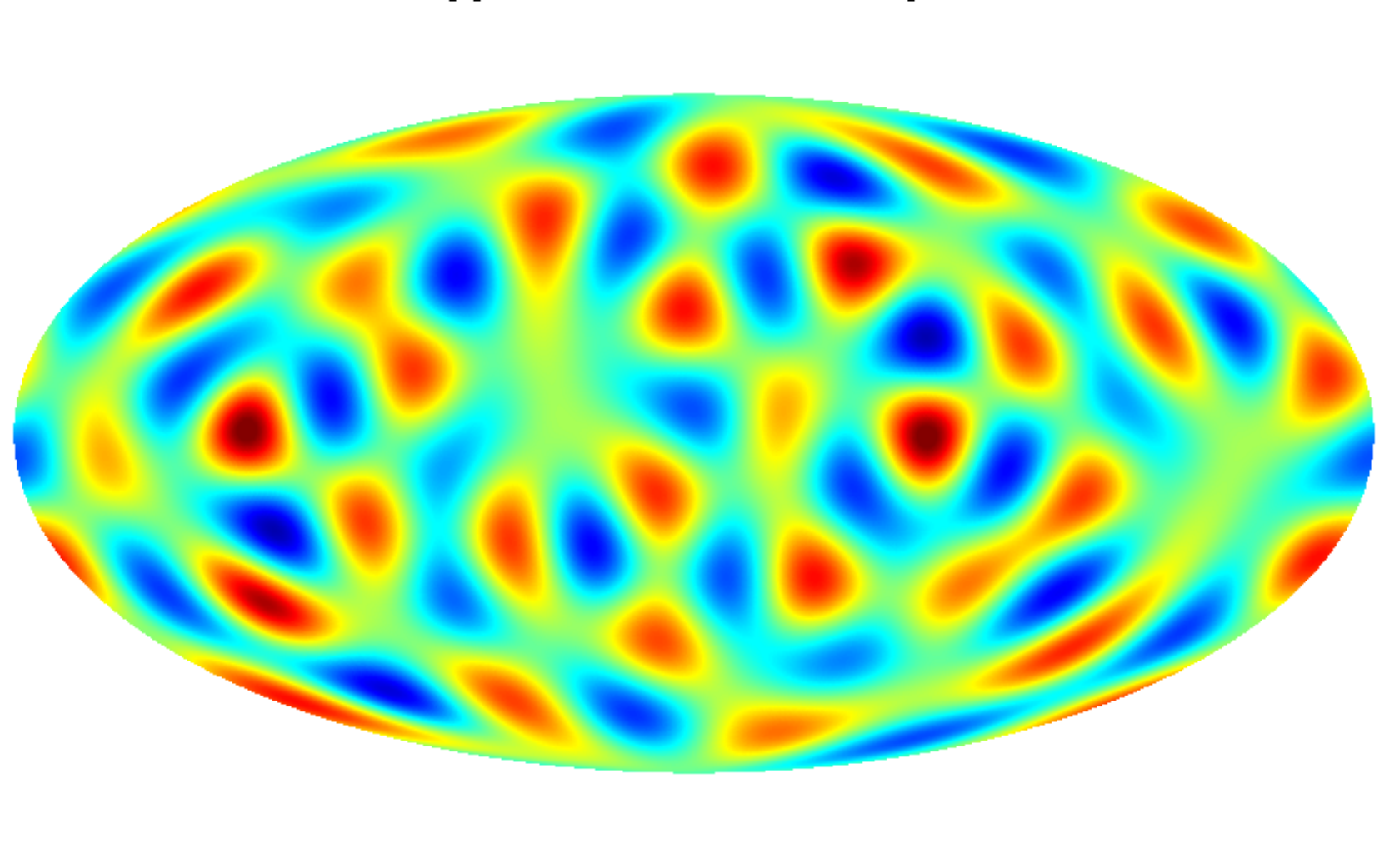} \\
\includegraphics[width=5cm]{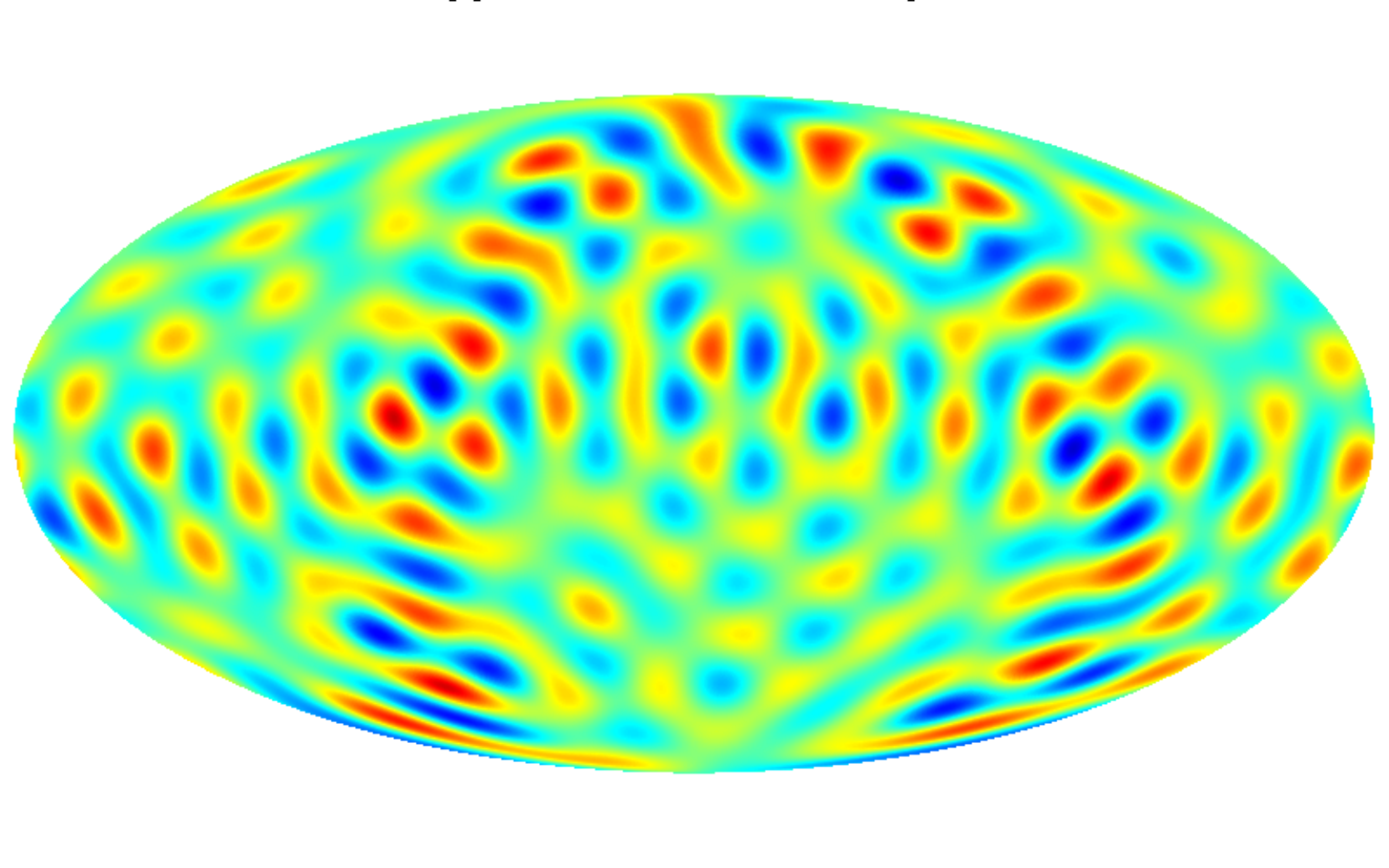}\includegraphics[width=5cm]{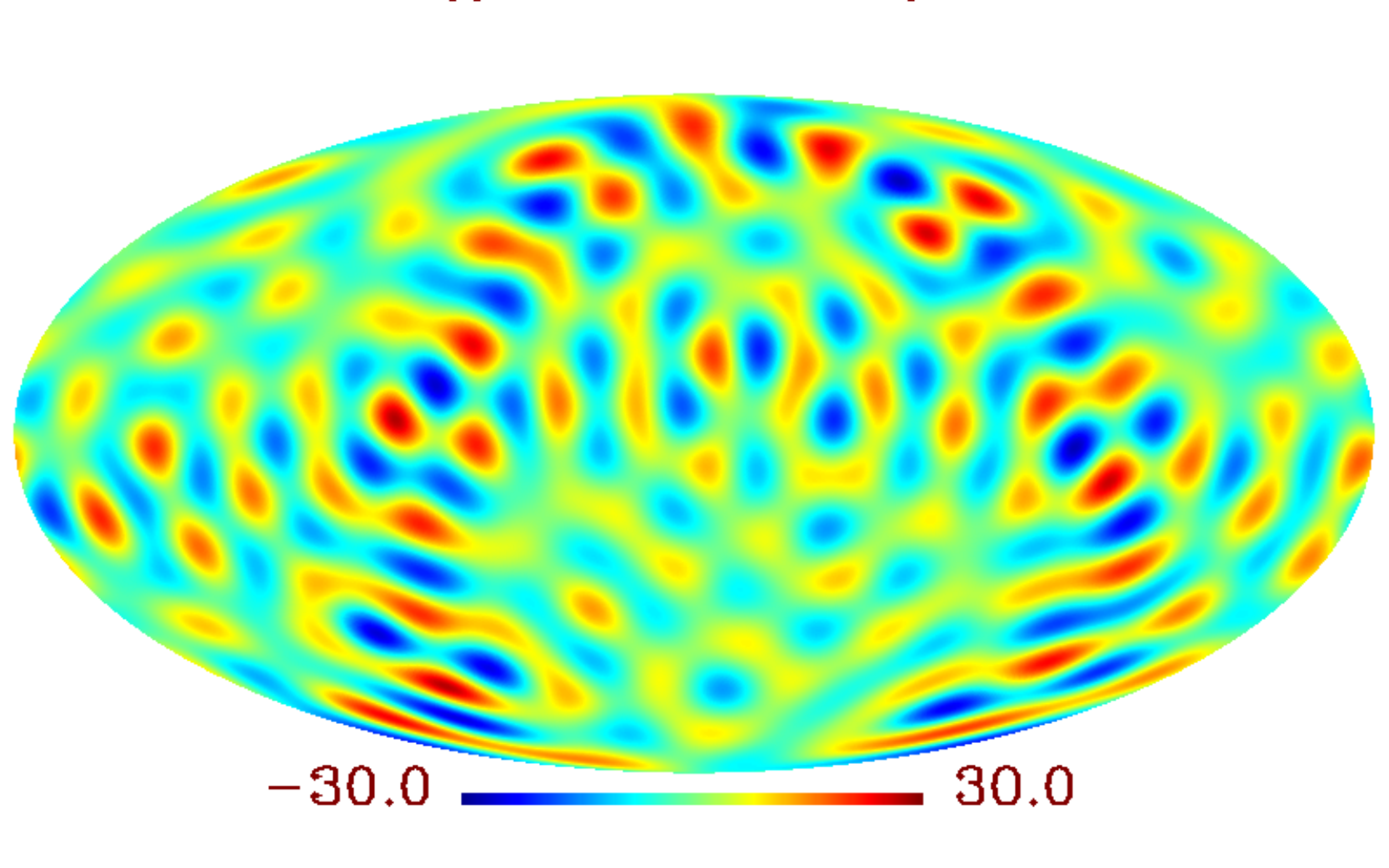}\includegraphics[width=5cm]{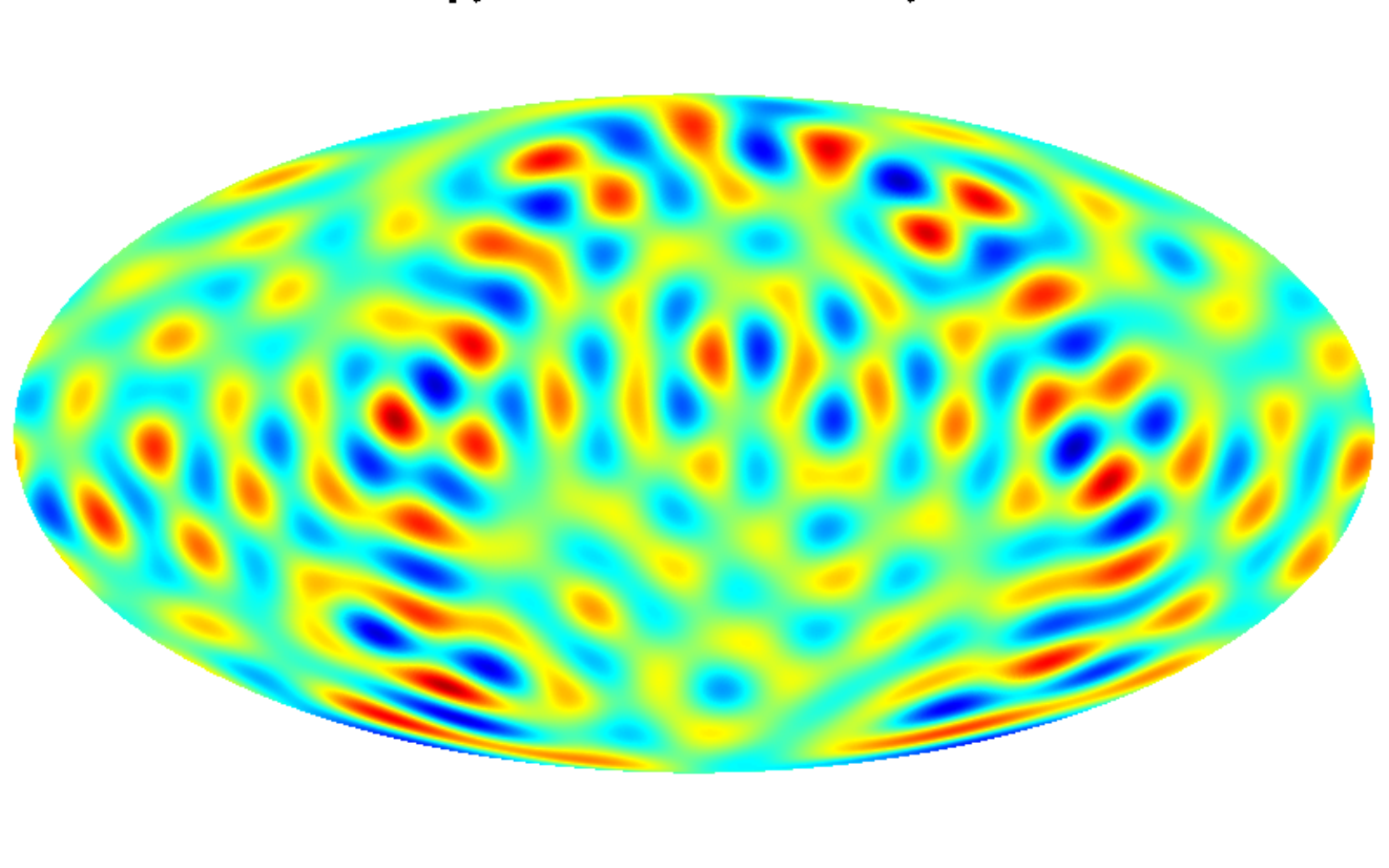}
\end{center}
\caption{The multipoles of $l=10$ (upper) and $l=17$ (lower). The left panels are derived from the NILC map, the middle ones are derived from
the SMICA map, and the right ones are from the SEVEM map. All panels use the
units $\mu$K.}\label{fig00}
\end{figure*}

{In the forthcoming discussion, we shall use the CMB low multipoles derived from the released Planck data, including
SMICA, NILC, and SEVEM. These CMB products are made via different techniques: the SMICA map is made via spectral parameter
fitting in the harmonic domain, NILC is made via a needlet variant of the internal linear combination technique, and SEVEM is constructed
through template fitting using the lowest- and highest-frequency bands. These maps have the obvious differences in the Galactic plane (see Ref. \cite{planckdata} for details). Here, we will demonstrate that in the low multiples the biases caused by the residual contaminations in these maps are relatively small as expected. In Fig. \ref{fig0}, we compare the power spectra from these full-sky maps with those derived from the masked Planck {\tc{Commander-Ruler (a pixel-based version of
parameter and template fitting)}} map \cite{planckdata}. We find that the differences between them are indeed small, especially for the multipoles $l<10$, which is consistent with the results in Ref. \cite{aluri}. When $l$ is larger, the biases become a little more obvious, especially for the multipoles $l=10$ and $l=17$. However, even for these two multipoles, we find the morphologies of them derived from three maps are nearly the same (see Fig. \ref{fig00}), which indicates that the residual contaminations in the maps have little effect on the morphologies of the CMB low multipoles.}

To study the directional properties of the CMB field,
similar to Ref. \cite{naselskyzhao}, we can define the rotationally variant
power spectrum $D(l)$ as
 \be\label{dl0}
 D_l\equiv \frac{1}{2l}\sum_{m=-l}^{l} |a_{lm}|^2(1-\delta_{m0}),
 \ee
where $\delta_{mm'}$ is the Kronecker symbol. Note that, the definition of $D_{l}$ in Eq. (\ref{dl0}) is slightly different from that defined in Ref. \cite{naselskyzhao}. For the random Gaussian, we find that $\langle D_{l}\rangle=C_{l}$; i.e., $D_{l}$ is the unbiased estimator for $C_{l}$, where $\langle ... \rangle$ denotes the average over the
statistical ensemble of realizations. Compared with the spectrum
in Eq. (\ref{cl}), in this definition, the $m=0$ component has been
excluded, so the $z$-axis direction has been selected as the
preferred direction in this definition.

Thus, we can define the power spectrum $D_l$ in any coordinate
system. Imagining the Galactic coordinate system is rotated by the
Euler angle $(0,\theta,\phi)$, and the coefficients $a_{lm}$ in
this new coordinate system are calculated by
 \be\label{alm_q}
 a_{lm}(\hat{\bf q})=\sum_{m'=-l}^{l} a_{lm'} D^{l}_{mm'}(0,\theta,\phi),
 \ee
where $\hat{\bf q}\equiv (\theta,\phi)$, $a_{lm}$ are
the coefficients defined in the Galactic coordinate system, and
$D^{l}_{mm'}(\psi,\theta,\phi)$ is the Wigner rotation matrix. If we consider  $\hat{\bf
q}$ as a vector, which labels the $z$-axis direction in the
rotated coordinate system, then $(\theta,\phi)$ is the polar
coordinate of this direction in the Galactic system. So
the general power spectrum $D_l(\hat{\bf q})$ is defined as
 \be\label{dl}
 D_l(\hat{\bf q})\equiv \frac{1}{2l}\sum_{m=-l}^{l} |a_{lm}(\hat{\bf q})|^2(1-\delta_{m0}).
 \ee

Now, we can define the parity statistic. As shown in Refs. \cite{kim,quadrupole}, for investigation of the parity asymmetry we can consider the statistic (see Table \ref{tab0})
 \be\label{g1}
 g_1(l,\hat{\bf q}) = \frac{\sum_{l'=2}^{l}l'(l'+1)D_{l'}(\hat{\bf q})\Gamma^{+}_{l'}}{\sum_{l'=2}^{l}l'(l'+1)D_{l'}(\hat{\bf q})\Gamma^{-}_{l'}},
 \ee
where $\Gamma^{+}_l=\cos^2\left(\frac{l\pi}{2}\right)$  and
$\Gamma^{-}_l=\sin^2\left(\frac{l\pi}{2}\right)$. This statistic
is associated with the degree of the parity asymmetry, where a
value of $g_1 < 1$ indicates the odd-parity preference, and $g_1 >
1$ indicates the even-parity preference. {For any given $l$, the sky map of $g_1(l,\hat{\bf q})$ can be constructed by considering all the directions $\hat{\bf q}$. In practice, we pixelize the full sky in the HEALPix format with the resolution parameter $N_{\rm side}=64$, and set the directions $\hat{\bf q}$ to be those of the pixels.}

For the temperature fluctuations of CMB map $\Delta
T(\theta,\phi)$, the two-point correlation function $C(\Theta)$
is naturally defined as
 \be
 C(\Theta,\hat{\bf q})=\sum_{l=2}^{\infty} \frac{2l+1}{4\pi} D_l(\hat{\bf q}) P_l(\cos\Theta),
 \ee
where $P_l(\cos\Theta)$ are the Legendre polynomials. Note that, in
this definition the contributions of $m=0$ components have been
excluded. For the largest angular distance $\Theta=\pi$, the
correlation function is
 \be
 C(\Theta=\pi,\hat{\bf q})=\sum_{l=2}^{\infty} \frac{2l+1}{4\pi} D_l(\hat{\bf q}) (\Gamma^{+}_l-\Gamma^{-}_l).
 \ee
So the natural way to estimate the relative contribution of even and odd multipoles to the correlation function is to define the statistic \cite{naselskyzhao} (see Table \ref{tab0}),
 \be\label{g2}
 g_2(l,\hat{\bf q}) = \frac{\sum_{l'=2}^{l}(2l'+1)D_{l'}(\hat{\bf q})\Gamma^{+}_{l'}}{\sum_{l'=2}^{l}(2l'+1)D_{l'}(\hat{\bf
 q})\Gamma^{-}_{l'}},
 \ee
which follows that $C(\Theta=\pi) \propto (g_2(l,\hat{\bf q})-1)$.
$g_2>1$ corresponds to the positive correlation of the opposite
directions, and $g_2<1$ indicates the anticorrelation of them.
Note that the statistic $g_2$ is different from $g_1$, due to the different factors before $D_l$ in their definitions.
Therefore, the relative weights of low multipoles are much higher in $g_2$ than those in $g_1$.

For further investigation, in this paper we also
consider the third statistic to quantify the parity asymmetry,
which was first introduced in Ref. \cite{aluri} (see Table \ref{tab0}),
 \be\label{g3}
 g_3(l,\hat{\bf q})= \frac{2}{l-1} \sum_{l'=3}^{l} \frac{(l'-1)l' D_{l'-1}(\hat{\bf q})}{l'(l'+1) D_{l'}(\hat{\bf q})},
 \ee
where the maximum, $l$, is any odd multipole $l\ge 3$ and the
summation is over all odd multipoles up to $l$. This statistic is the
measure of the mean deviation of the ratio of power in an even
multipole to its succeeding odd-multipole form one.

\begin{figure*}[t]
\begin{center}
\includegraphics[width=5cm]{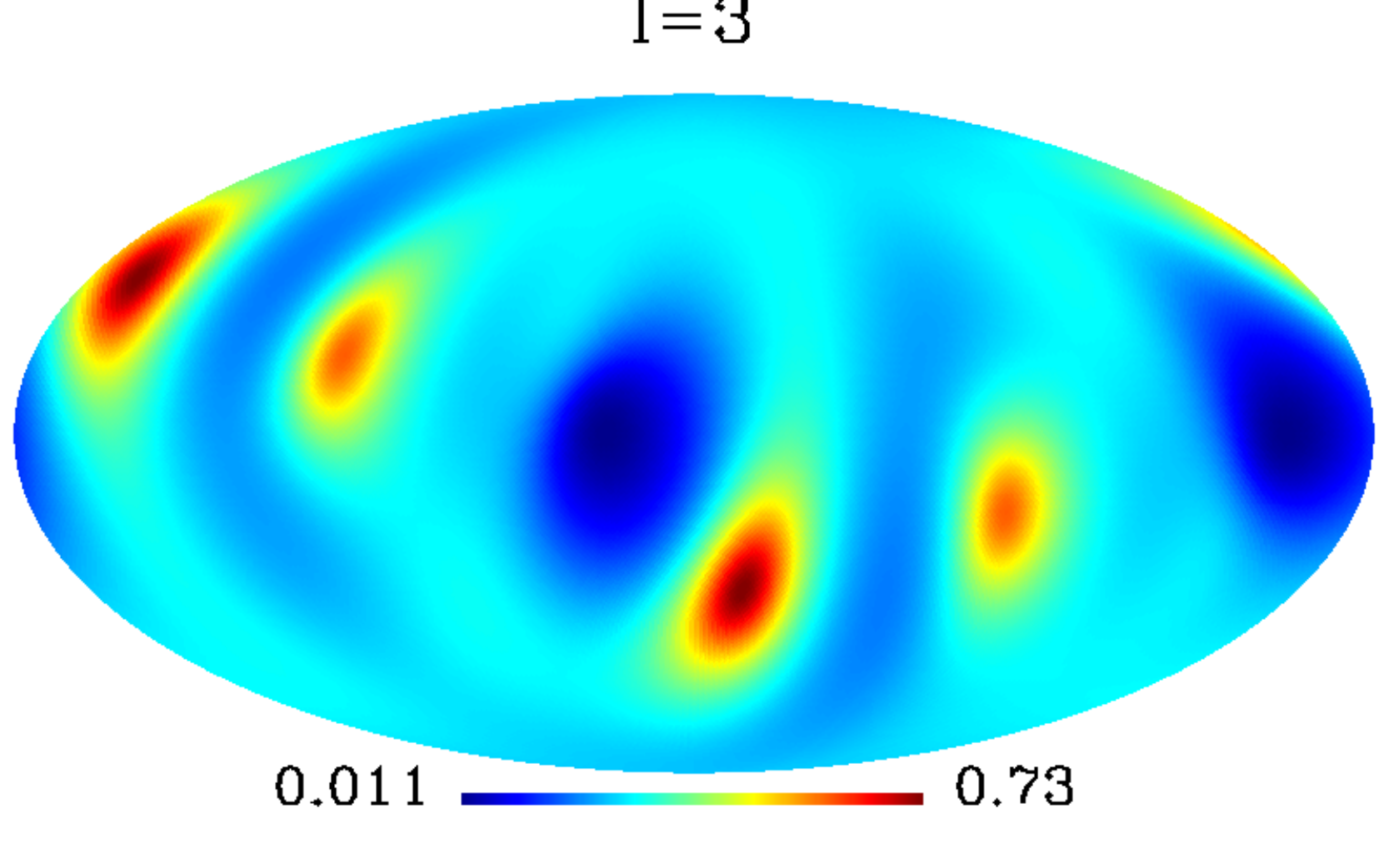}\includegraphics[width=5cm]{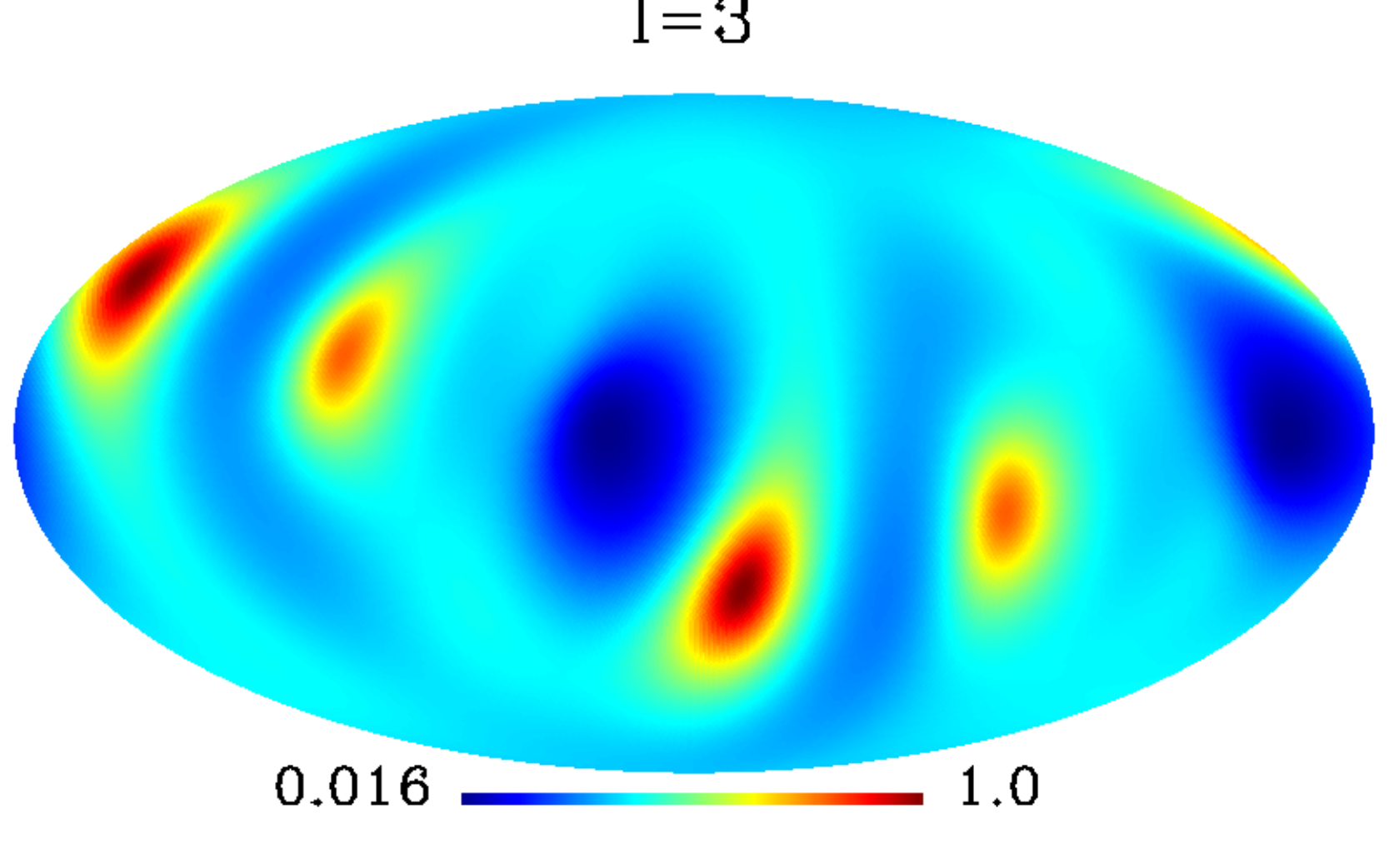}\includegraphics[width=5cm]{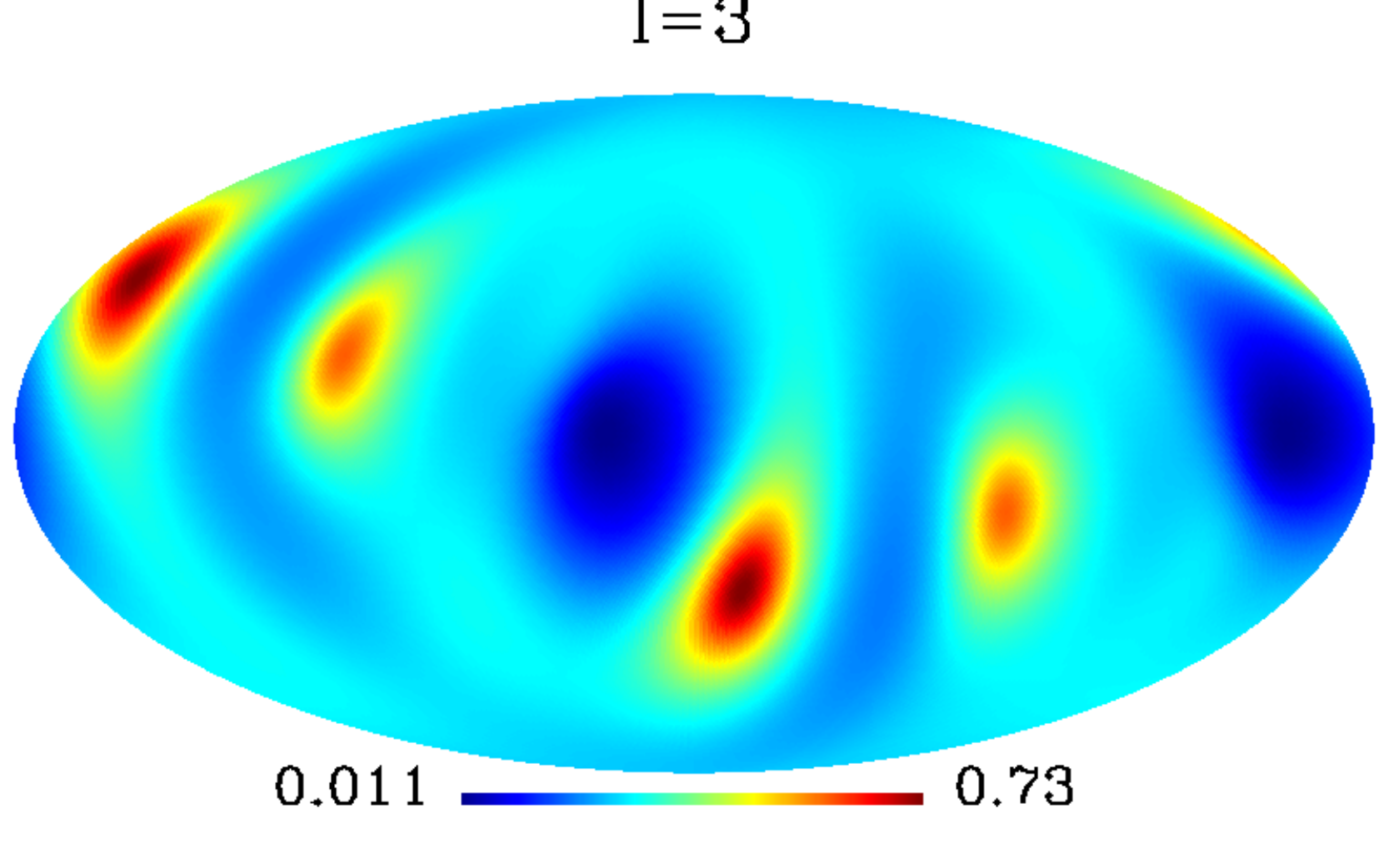} \\
\includegraphics[width=5cm]{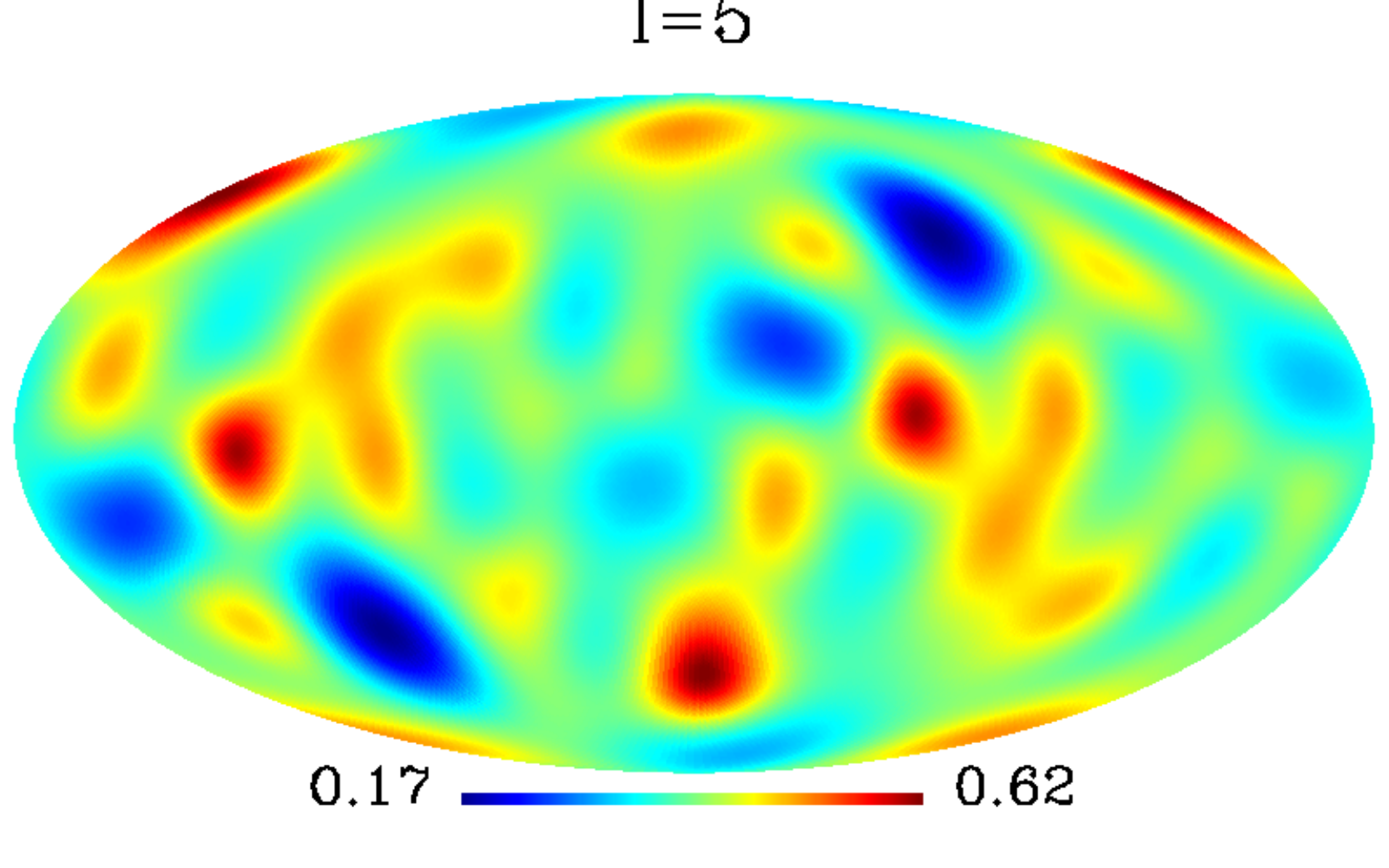}\includegraphics[width=5cm]{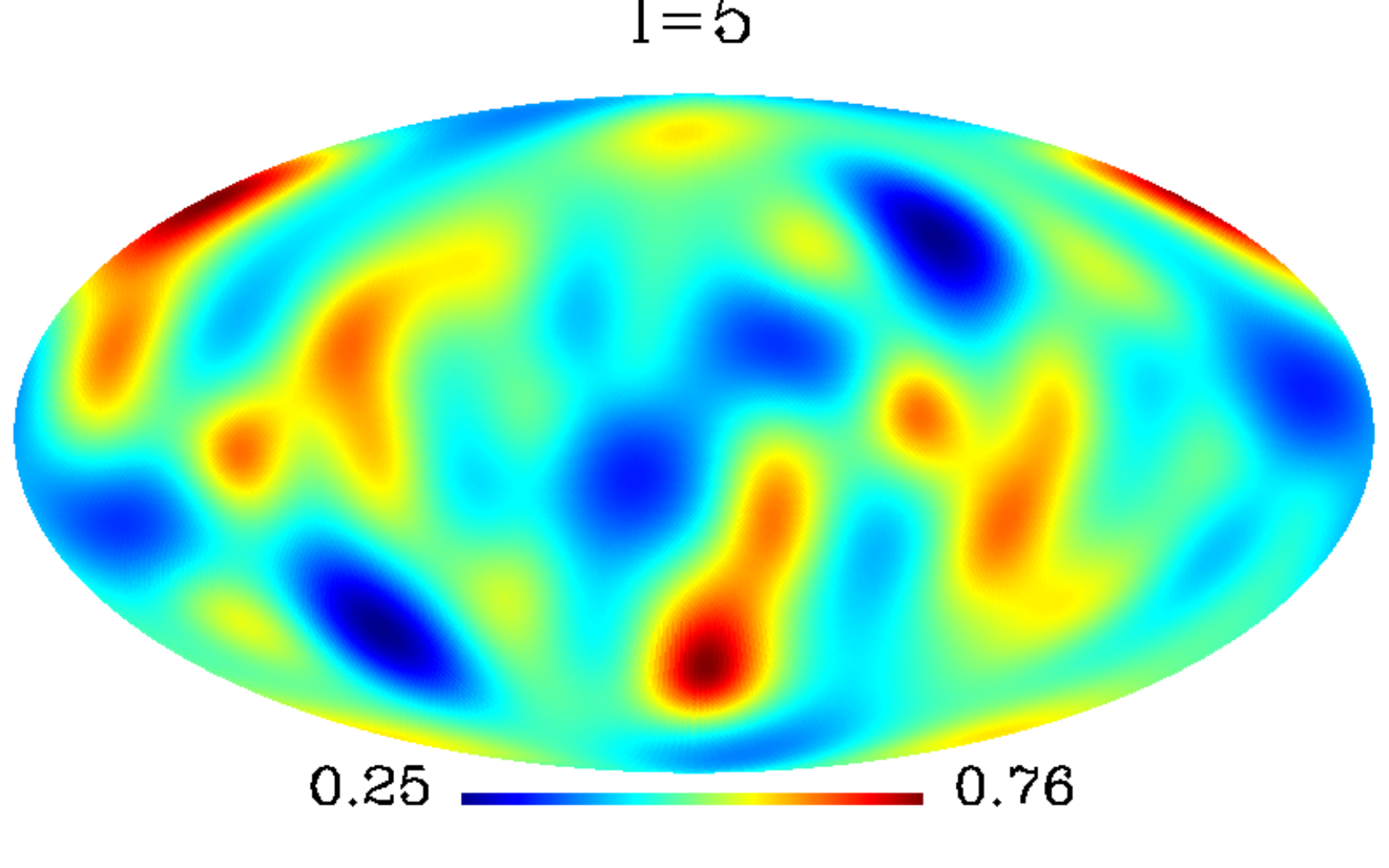}\includegraphics[width=5cm]{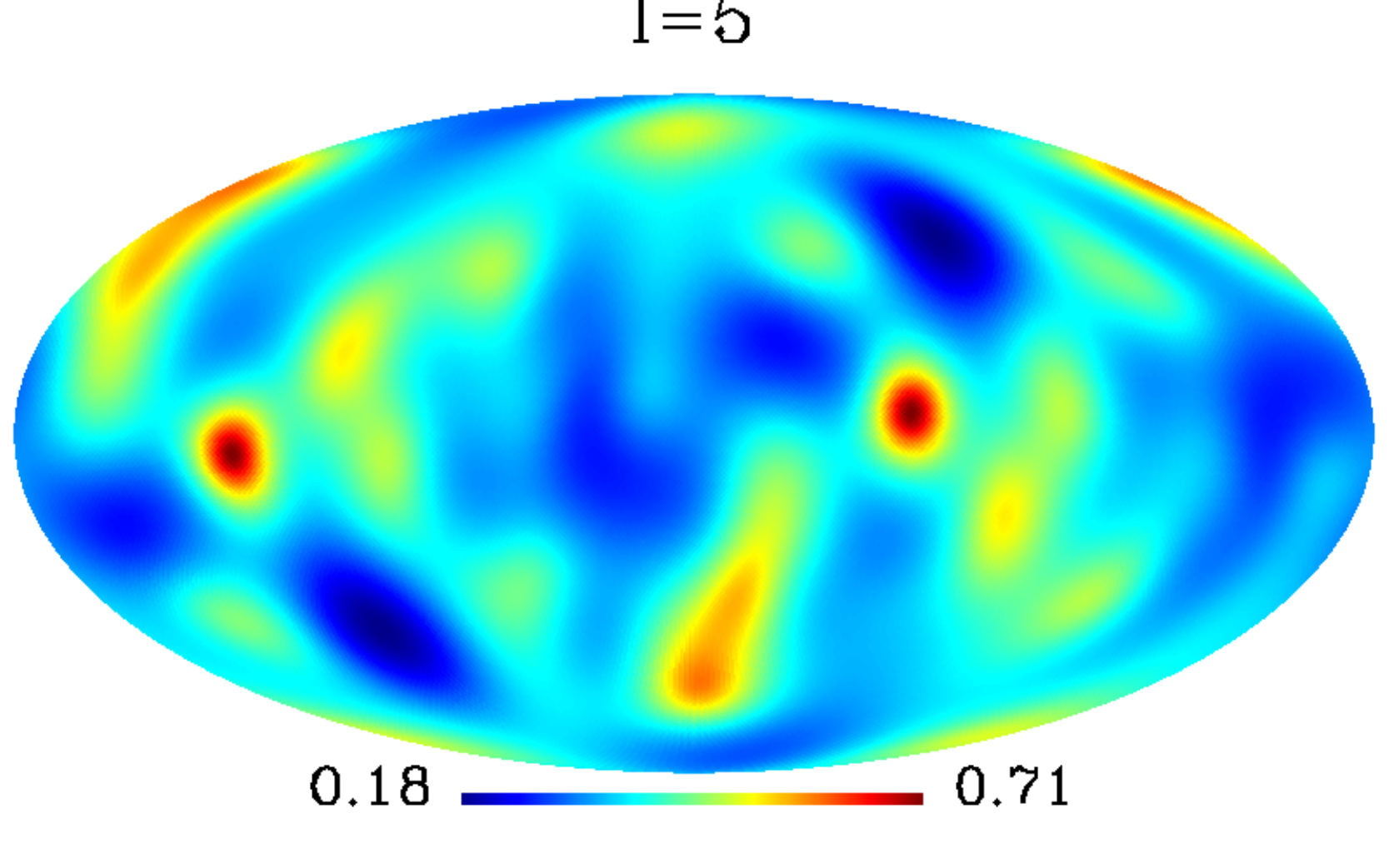} \\
\includegraphics[width=5cm]{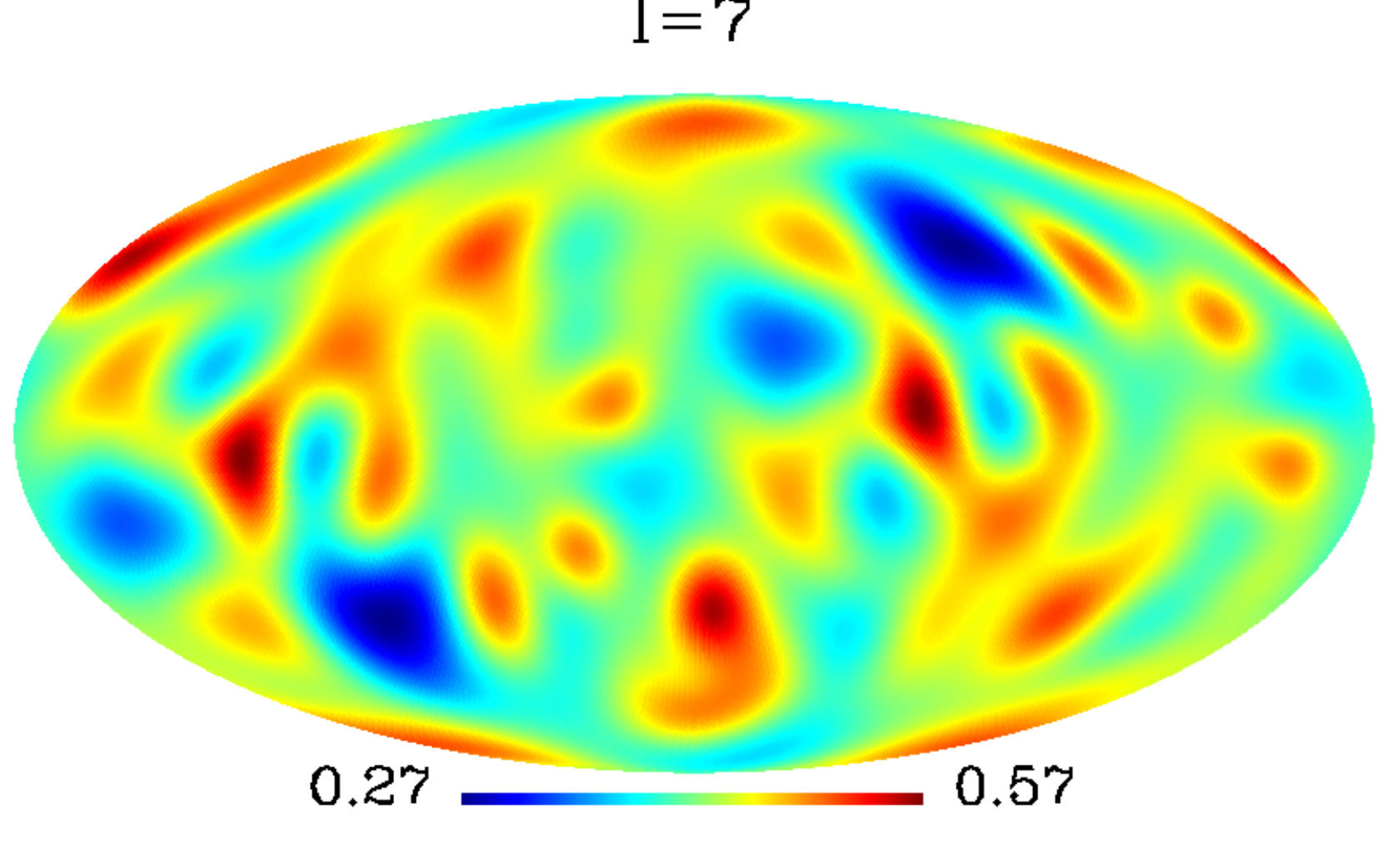}\includegraphics[width=5cm]{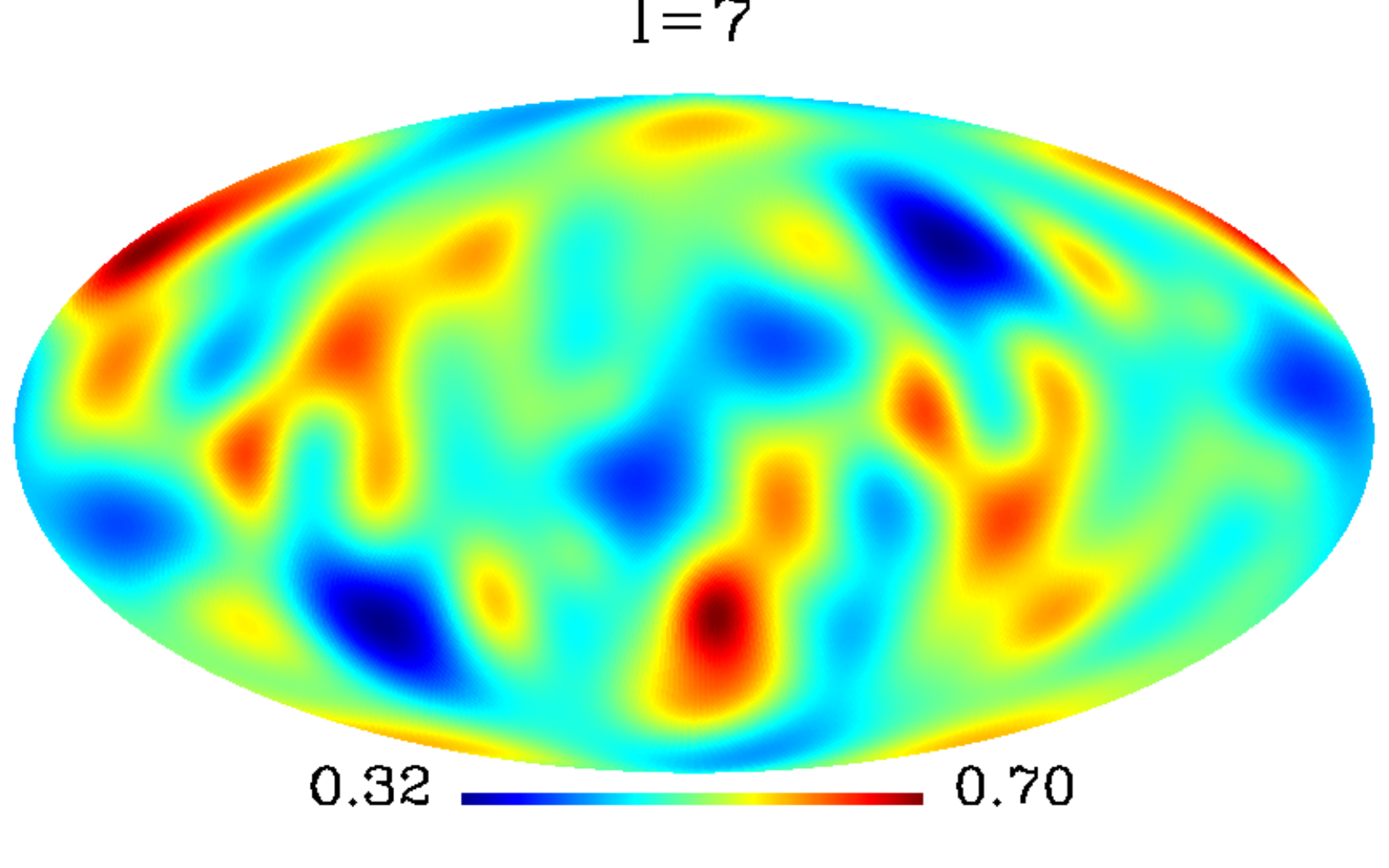}\includegraphics[width=5cm]{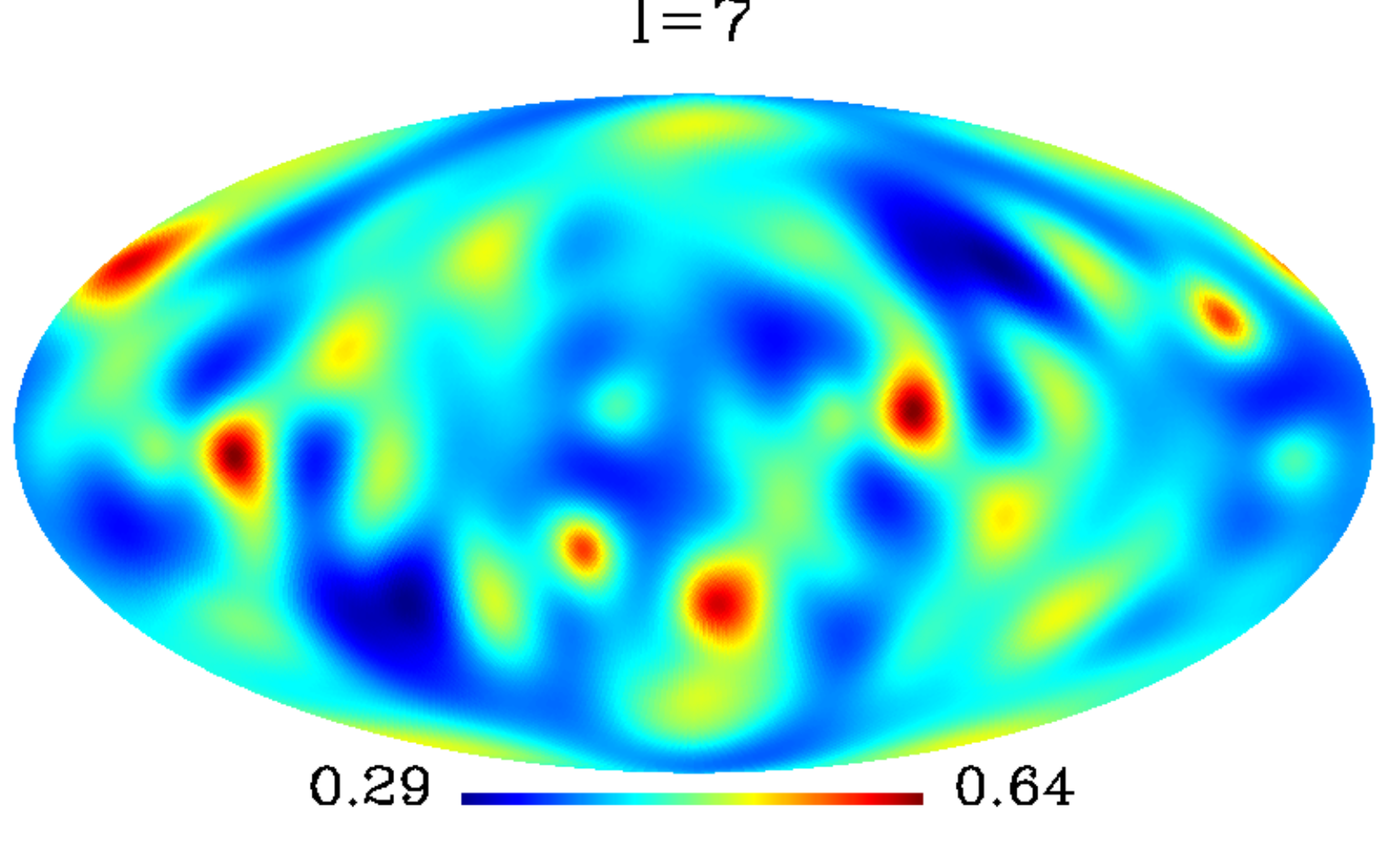} \\
\includegraphics[width=5cm]{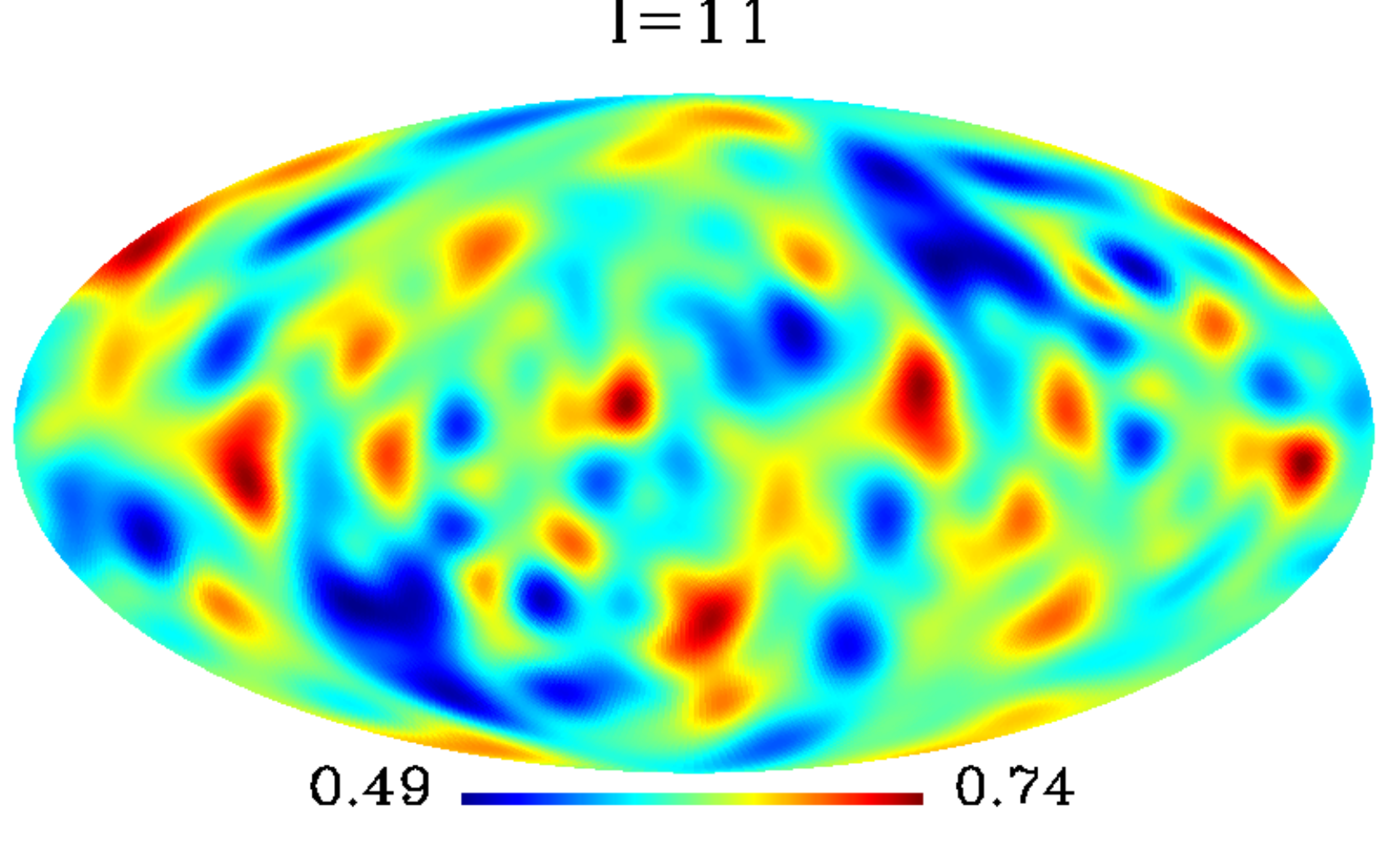}\includegraphics[width=5cm]{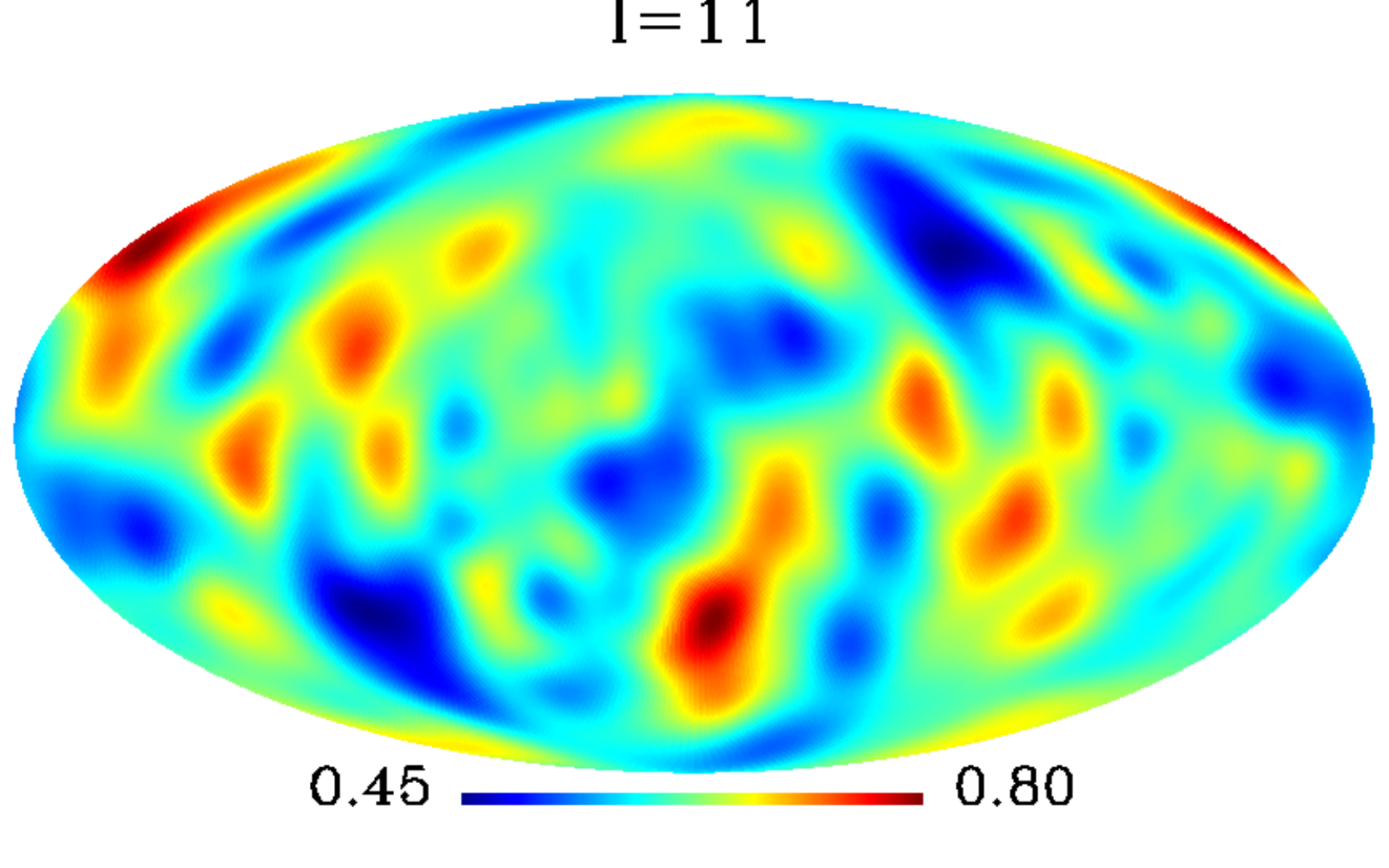}\includegraphics[width=5cm]{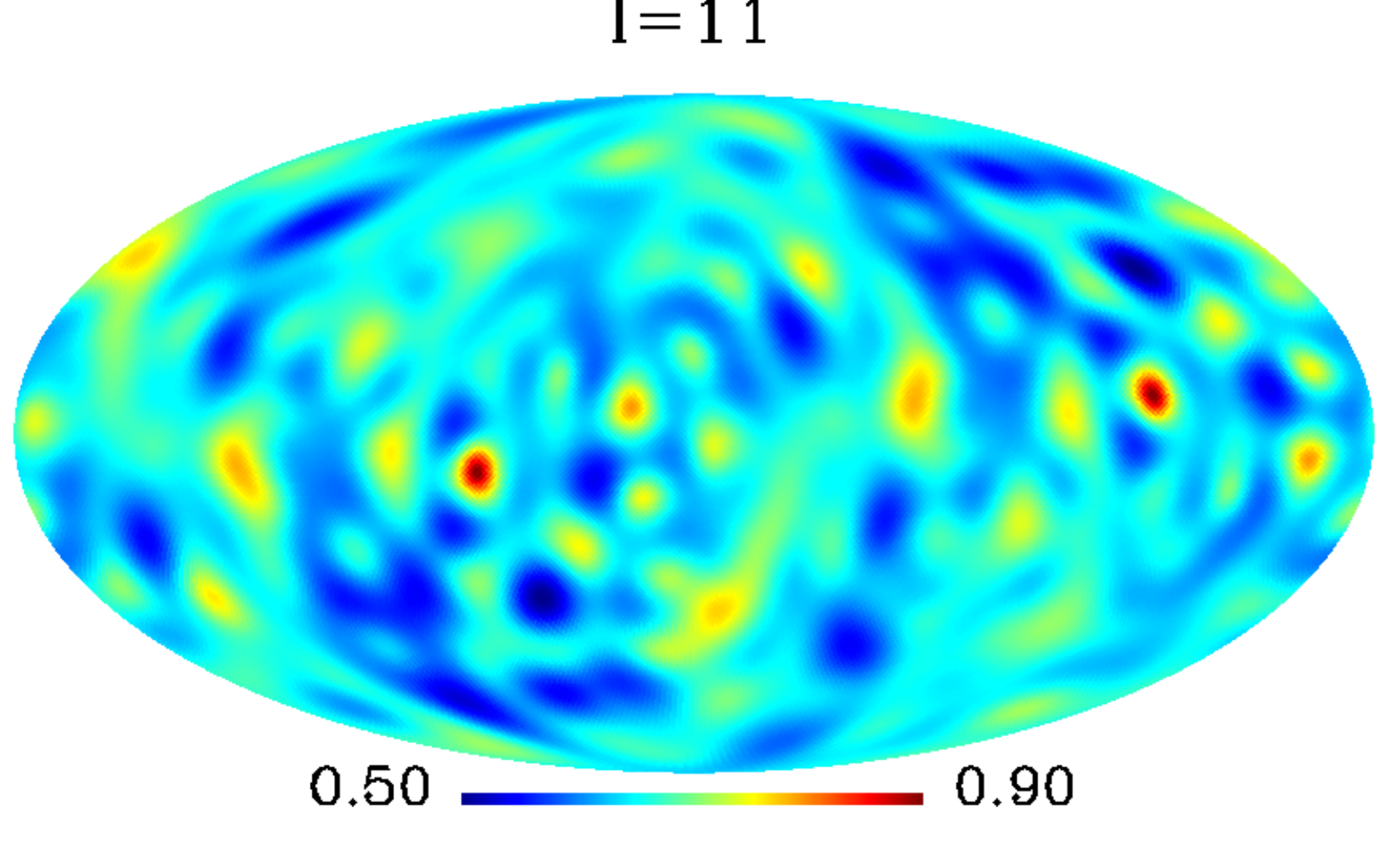} \\
\includegraphics[width=5cm]{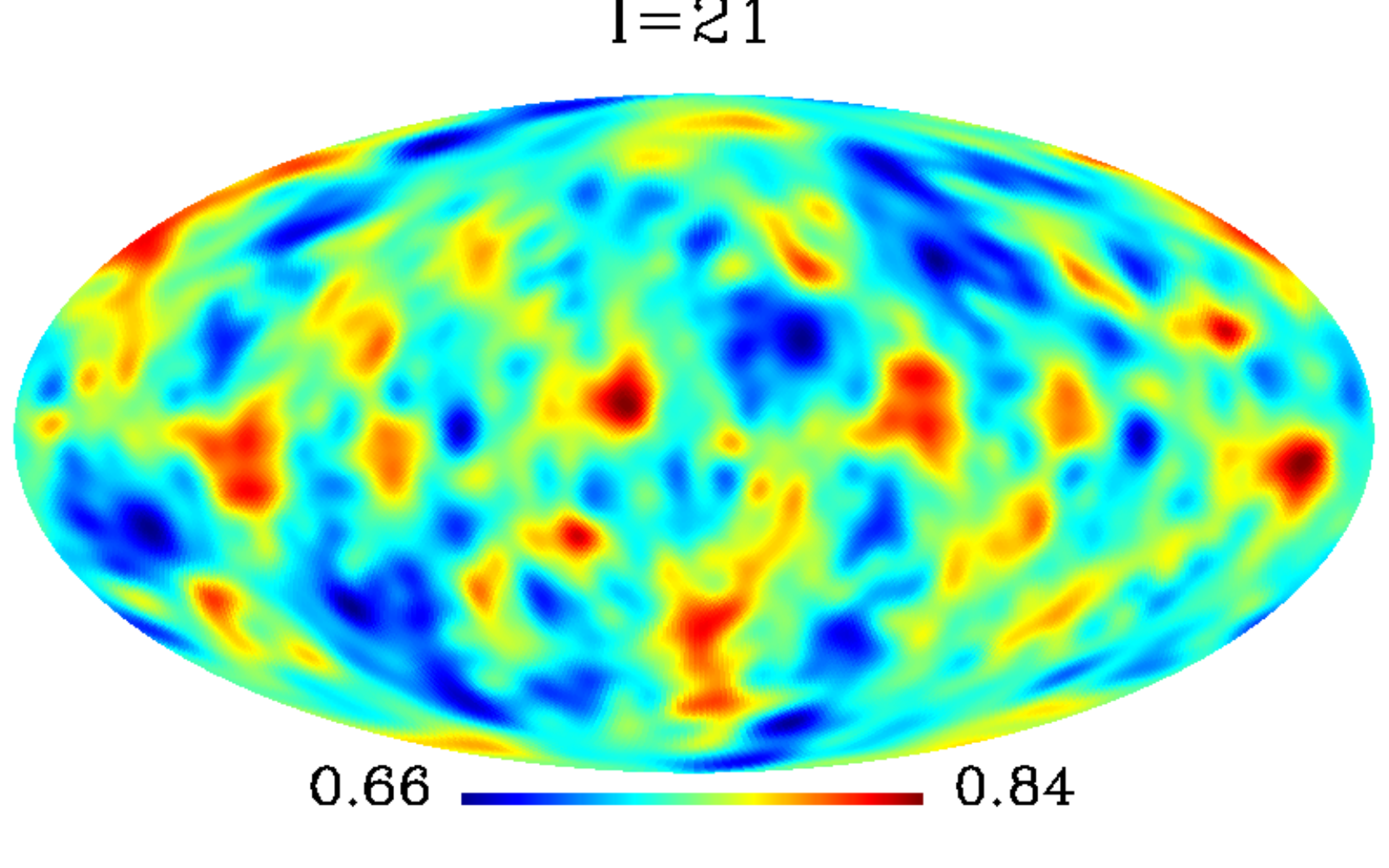}\includegraphics[width=5cm]{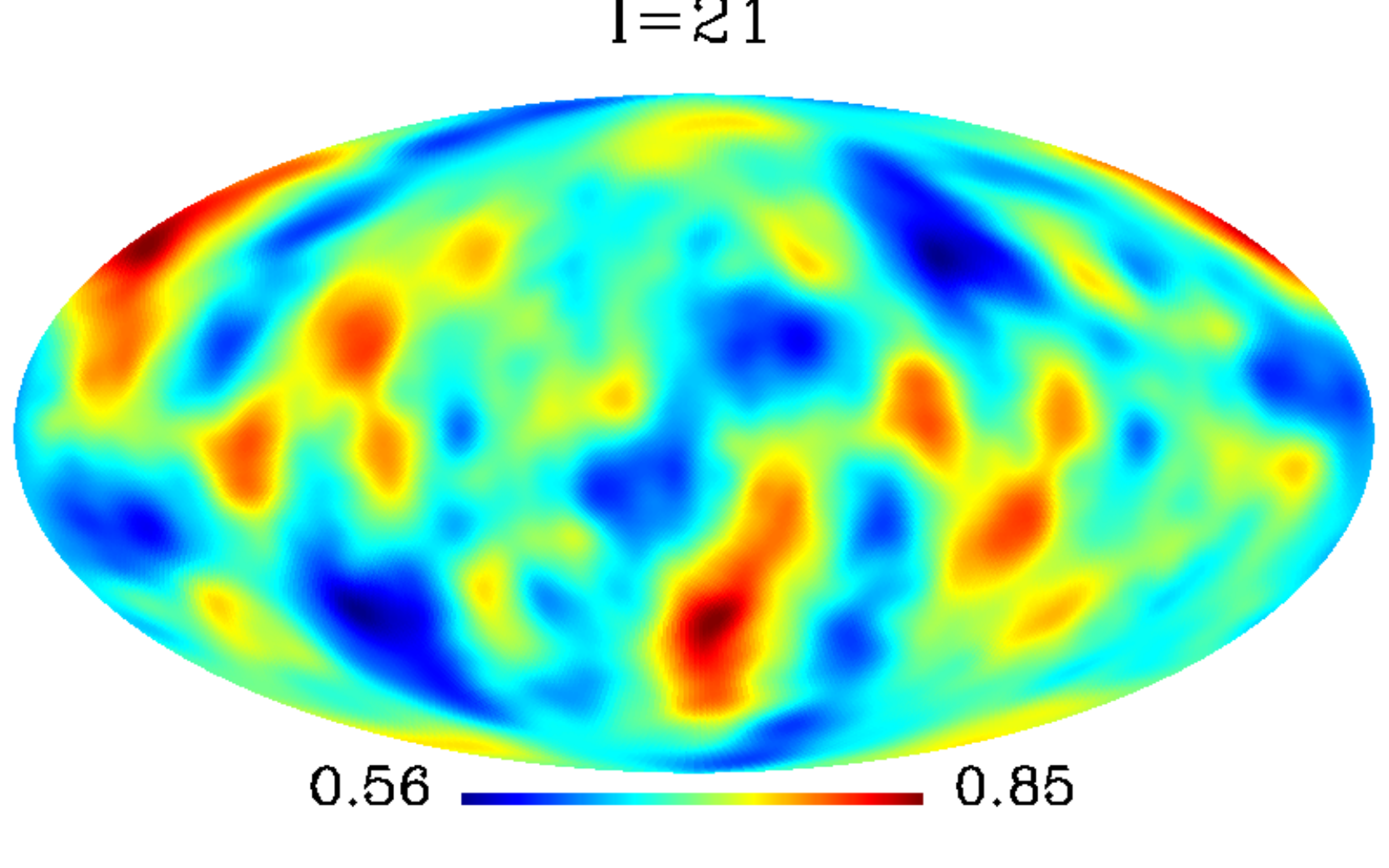}\includegraphics[width=5cm]{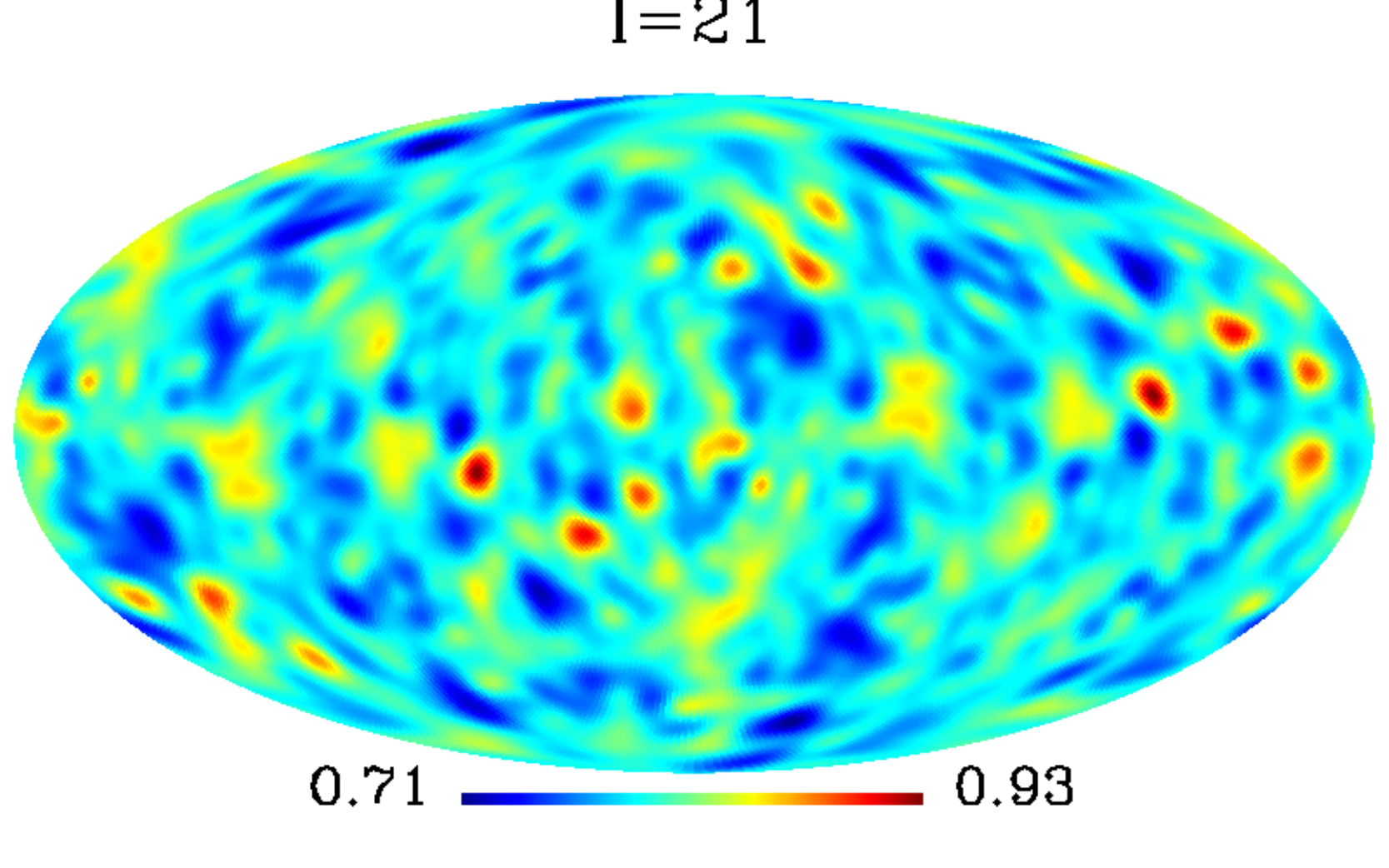}
\end{center}
\caption{Three directional statistics $g_1(l,\hat{\bf q})$ (left),
$g_2(l,\hat{\bf q})$ (middle), and $g_3(l,\hat{\bf q})$ (right) as
functions of $\hat{\bf q}\equiv (\theta,\phi)$. Note that,
these results are based on the Planck SMICA data.}\label{fig1}
\end{figure*}

{In previous works \cite{kim,kim2,aluri,quadrupole}, by using the power spectrum $C_l$,
the authors found that CMB parity asymmetry is quite significant at the low multipoles, and this tendency extends to the multipole range $l<22$. Since the definition of the estimator $D_l$ is similar to that of $C_l$, one expects the parity asymmetry of the statistics $g_1$, $g_2$, $g_3$ to also extend to this multipole range.}
We apply these three statistics for all the odd multipoles $3\le l
\le 21$ to the released Planck SMICA, NILC and SEVEM data. The results for SMICA
data are presented in Fig. \ref{fig1}. For all the odd maximum multipoles $l$
and directions $\hat{\bf q}$, we have $g_i<1$ for $i=1,2,3$. These
are also correct for Planck NILC and SEVEM data. So consistent
with previous works \cite{kim,kim2,naselskyzhao,aluri,quadrupole}, we
find that the real CMB data have the odd-parity preference, which
is independent of the choice of the parity statistics.

\begin{table}
\caption{The definitions of six directional statistics considered in the text.}
\begin{center}
\label{tab0}
\begin{tabular}{ |c|c|  }
   \hline
   Number of statistic  & Definition \\
   \hline
   $1^{\rm st}$  &    $g_1(l,\hat{\bf q})$ with $D_l(\hat{\bf q})$       \\
      \hline
   $2^{\rm nd}$  &    $g_2(l,\hat{\bf q})$ with $D_l(\hat{\bf q})$       \\
      \hline
   $3^{\rm rd}$  &    $g_3(l,\hat{\bf q})$ with $D_l(\hat{\bf q})$       \\
      \hline
   $4^{\rm th}$  &    $g_1(l,\hat{\bf q})$ with $\tilde{D}_l(\hat{\bf q})$       \\
      \hline
   $5^{\rm th}$  &    $g_2(l,\hat{\bf q})$ with $\tilde{D}_l(\hat{\bf q})$       \\
      \hline
   $6^{\rm th}$  &    $g_3(l,\hat{\bf q})$ with $\tilde{D}_l(\hat{\bf q})$       \\
   \hline
\end{tabular}
\end{center}
\end{table}

To cross-check the results, we consider another
rotationally variant estimator, proposed in Refs. \cite{tegmark,evil,naselskyzhao,quadrupole},
 \be\label{dl_tilde}
 \tilde{D}_{l} \equiv \frac{1}{2l+1} \sum_{m=-l}^{l} m^2 |a_{lm}|^2.
 \ee
As well discussed, this statistic has also selected the $z$-axis
direction as the preferred direction.
{However, we should mention that this estimator, $\tilde{D}_l$, is definitely different from $D_l$ defined in Eq. (\ref{dl0}).
First, for each multipole $l$, the components $a_{lm}$ with $m\neq0$ have the exact same weights in the definition of $D_l$,
which shows that this statistic only favors the component $m=0$. But from Eq. (\ref{dl_tilde}), we know that $\tilde{D}_l$
favors the high $m$s and so it works well in searches for planarity (i.e., $m=\pm l$) \cite{tegmark,evil}.
Secondly, comparing the different multipoles, we find that due to the factor $m^2$ in the definition,
the value of $\tilde{D}_{l}$ increases dramatically with the increasing of the multipole number $l$, which is also different from that of $D_l$.}

Because of the rotational
variance of this quantity, we can define the general power
spectrum as
 \be\label{tilde_dl}
 \tilde{D}_l(\hat{\bf q})\equiv \frac{1}{2l+1}\sum_{m=-l}^{l} m^2|a_{lm}(\hat{\bf q})|^2,
 \ee
where $a_{lm}(\hat{\bf q})$ is defined in Eq. (\ref{alm_q}). Thus, the other three directional statistics can be defined as follows (see also Table \ref{tab0}):
 \beqa\label{g456}
 g_4(l,\hat{\bf q})&=&\left.g_1(l,\hat{\bf q})\right|_{{D}_{l} \rightarrow\tilde{D}_{l}}, \\
 g_5(l,\hat{\bf q})&=&\left.g_2(l,\hat{\bf q})\right|_{{D}_{l} \rightarrow\tilde{D}_{l}}, \\
 g_6(l,\hat{\bf q})&=&\left.g_3(l,\hat{\bf q})\right|_{{D}_{l} \rightarrow\tilde{D}_{l}}.
 \eeqa

{As we have emphasized, the estimator $\tilde{D}_l$ is quite different from $D_l$ due to the factor $m^2$ in the definition. In the statistics $g_4$, $g_5$, or $g_6$, the contributions of the higher multipoles, $l\sim l_{\max}$, become completely dominant. For this reason, we only apply these statistics to the multipole range in which the CMB parity asymmetry is most obvious. Although, in Refs. \cite{naselskyzhao,quadrupole}, the authors found the CMB parity asymmetry can extend to multipole ranges up to $l \sim 22$, the main contribution comes from the lowest multipoles, i.e., $l<10$, which can be clearly seen in the following facts.
From Fig. \ref{fig0}, we find that the regular oscillation of the power spectra (i.e., the values of even multipoles are significantly smaller, while those of odd multipoles are significantly larger, which is the performance of the CMB parity asymmetry) is obvious only in the multipole range $l<10$. Consistently, from Fig. 1 in Refs. \cite{naselskyzhao}, we also find that the CMB parity asymmetry becomes negligible if the low multipoles $l<10$ are excluded.
So, the same as in Ref. \cite{naselskyzhao}, in this paper we only consider the parity statistics $g_4$, $g_5$, and $g_6$ for the multipoles $l < 10$.}

\begin{figure*}[t]
\begin{center}
\includegraphics[width=5cm]{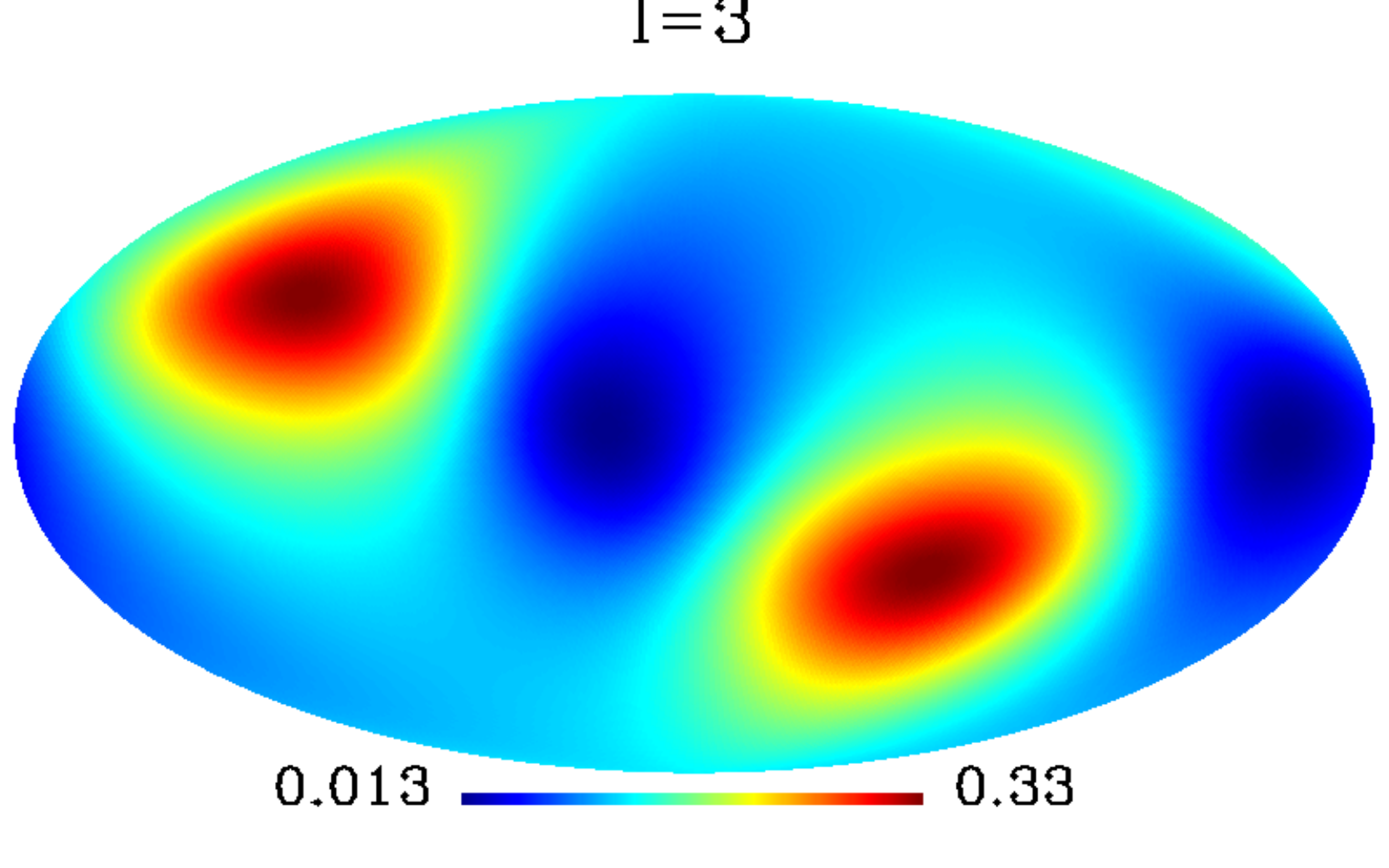}\includegraphics[width=5cm]{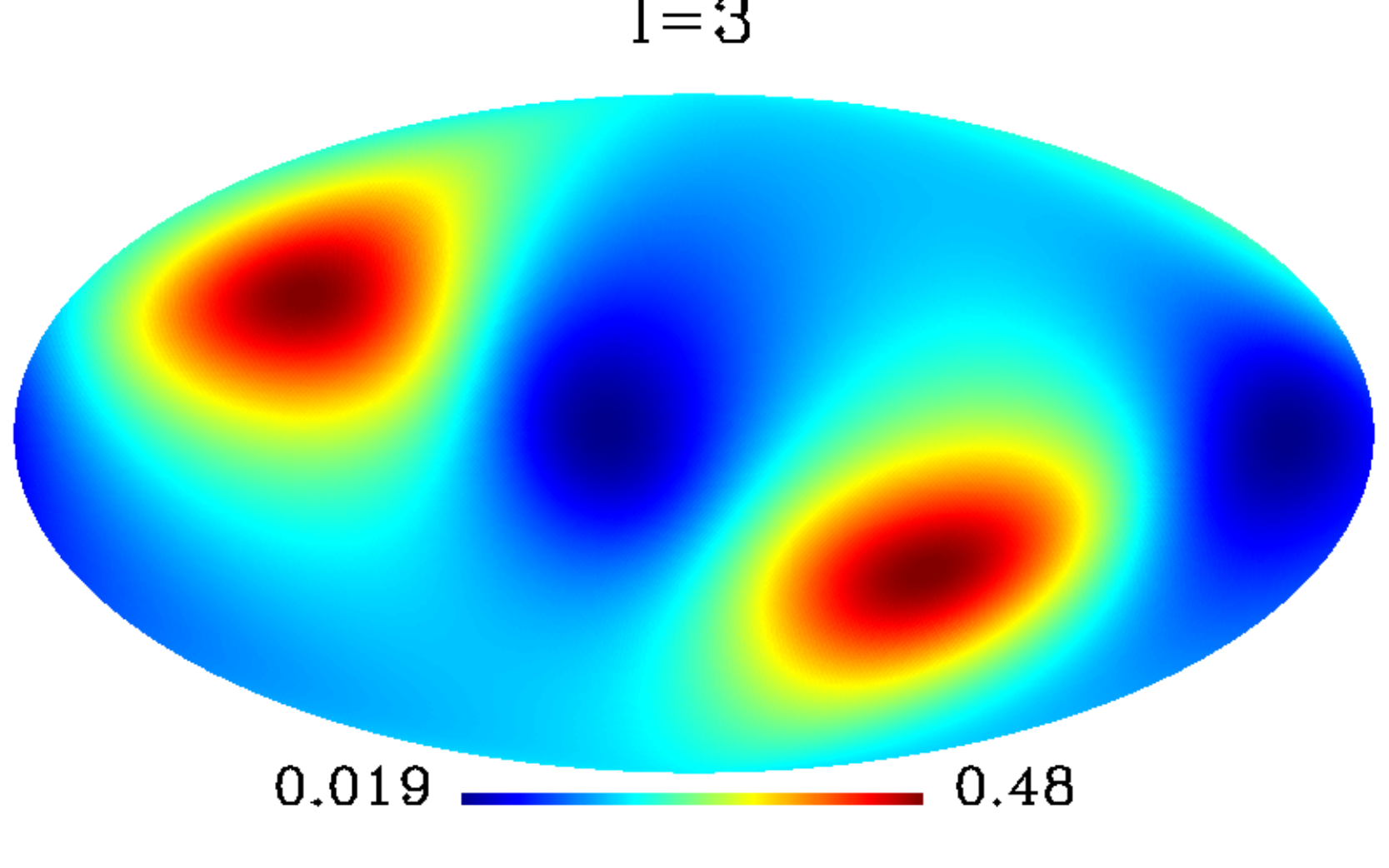}\includegraphics[width=5cm]{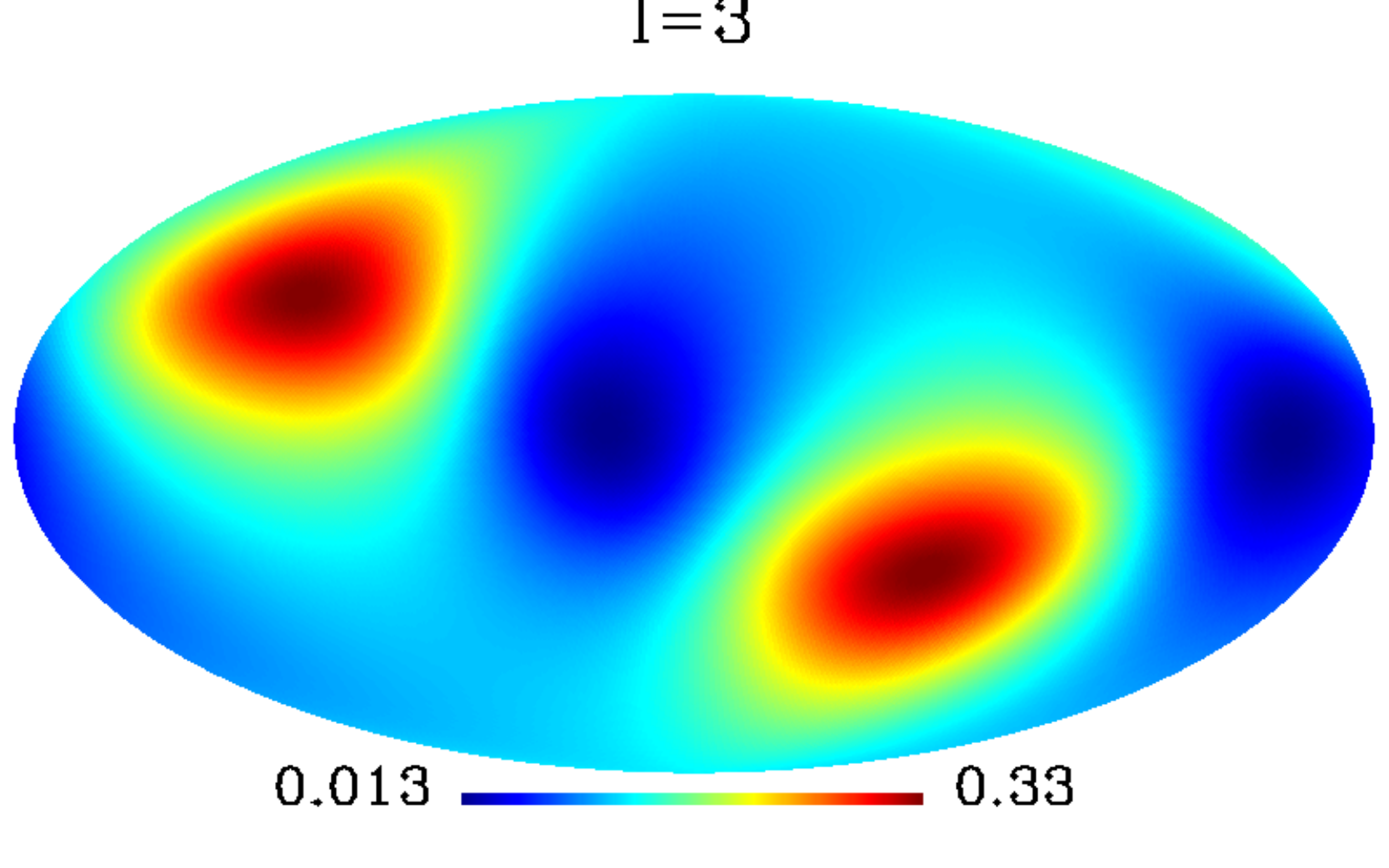} \\
\includegraphics[width=5cm]{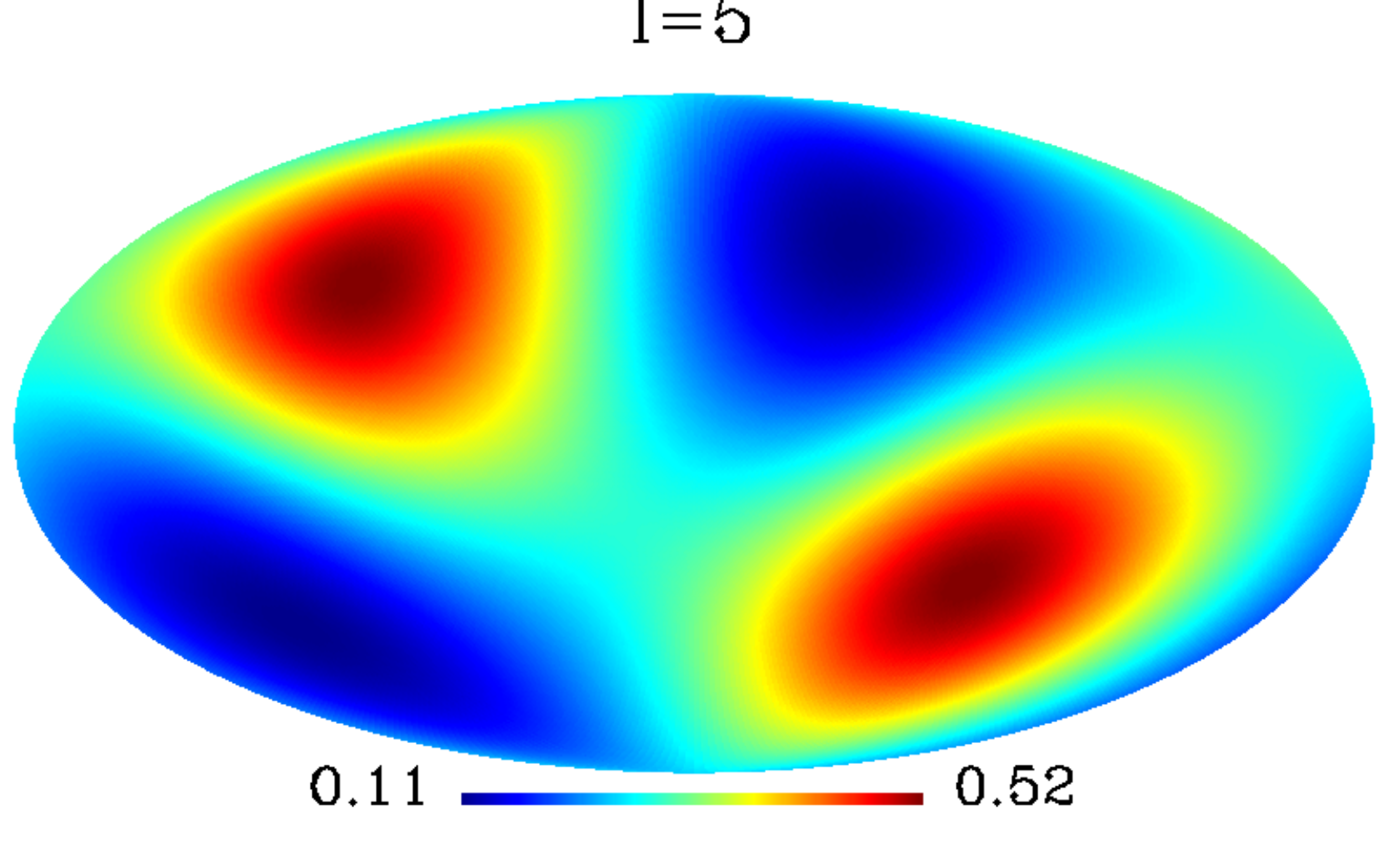}\includegraphics[width=5cm]{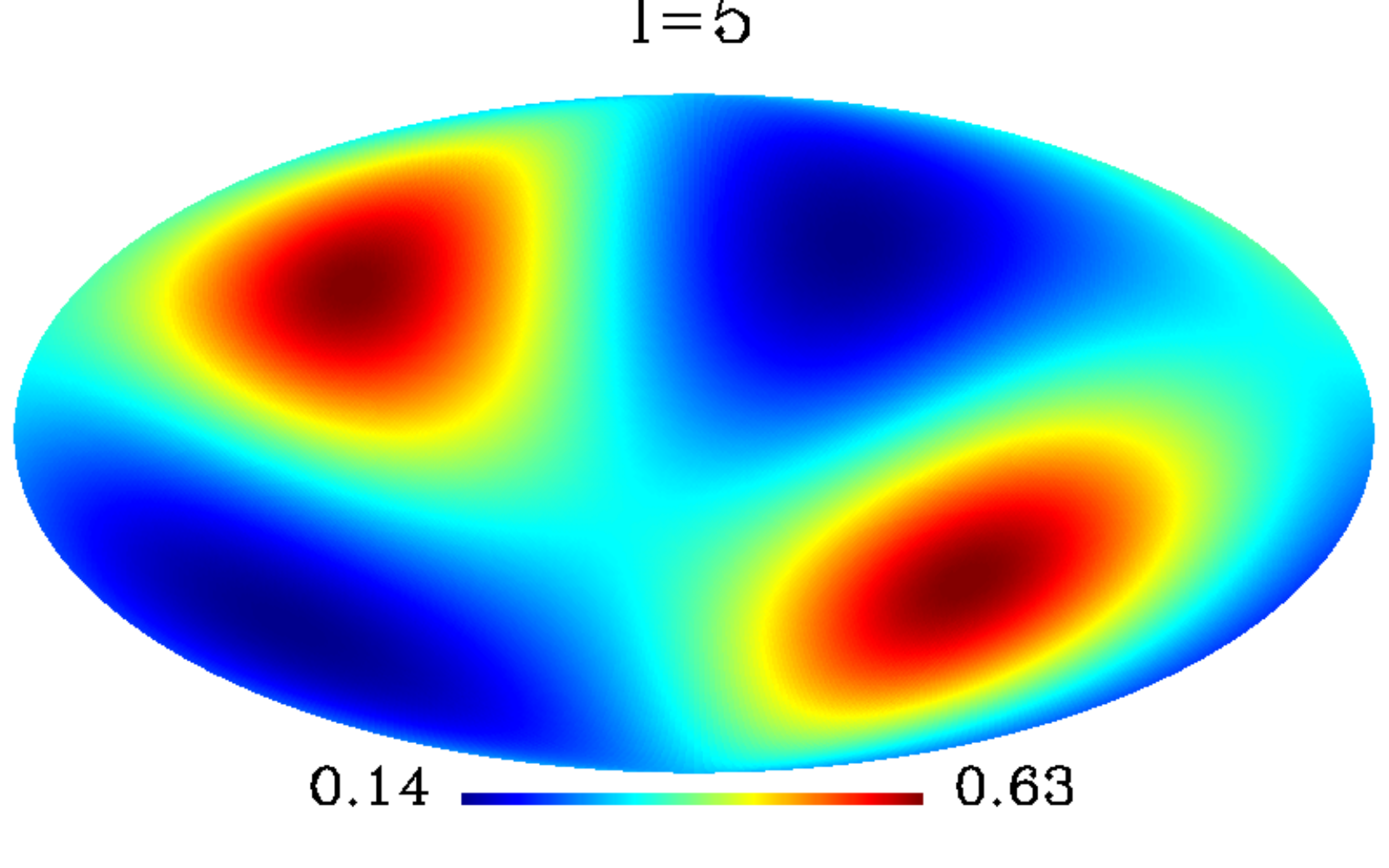}\includegraphics[width=5cm]{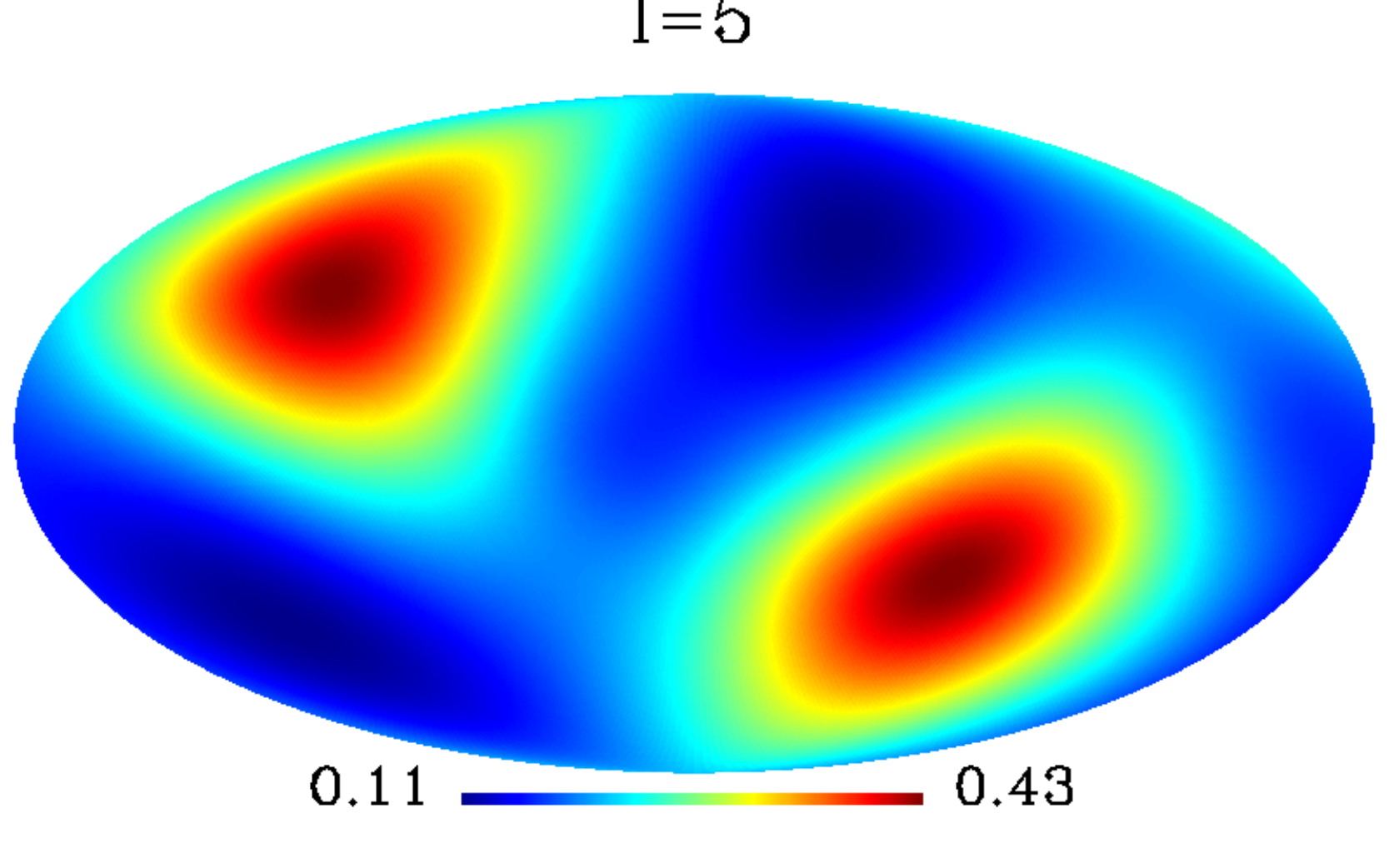} \\
\includegraphics[width=5cm]{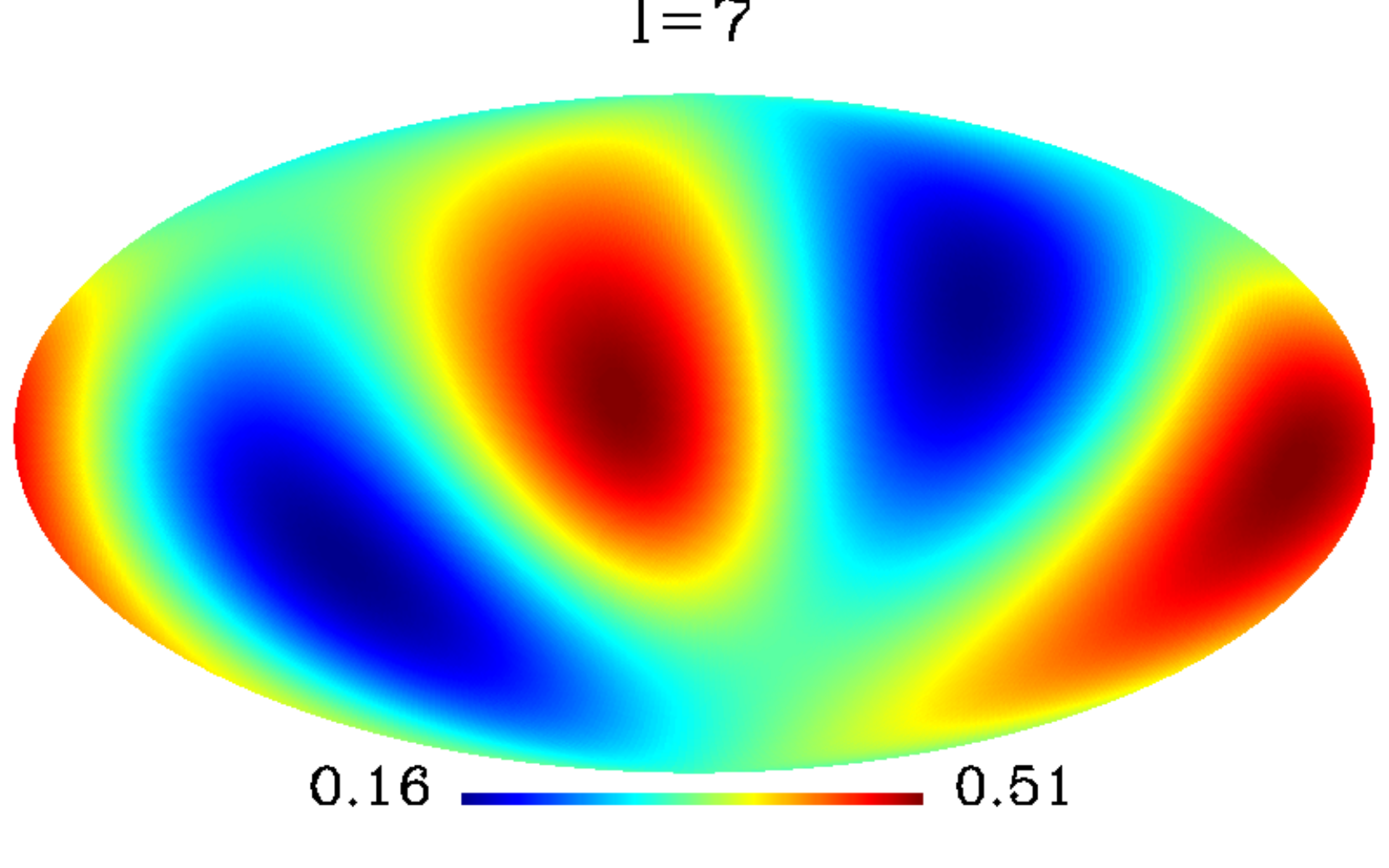}\includegraphics[width=5cm]{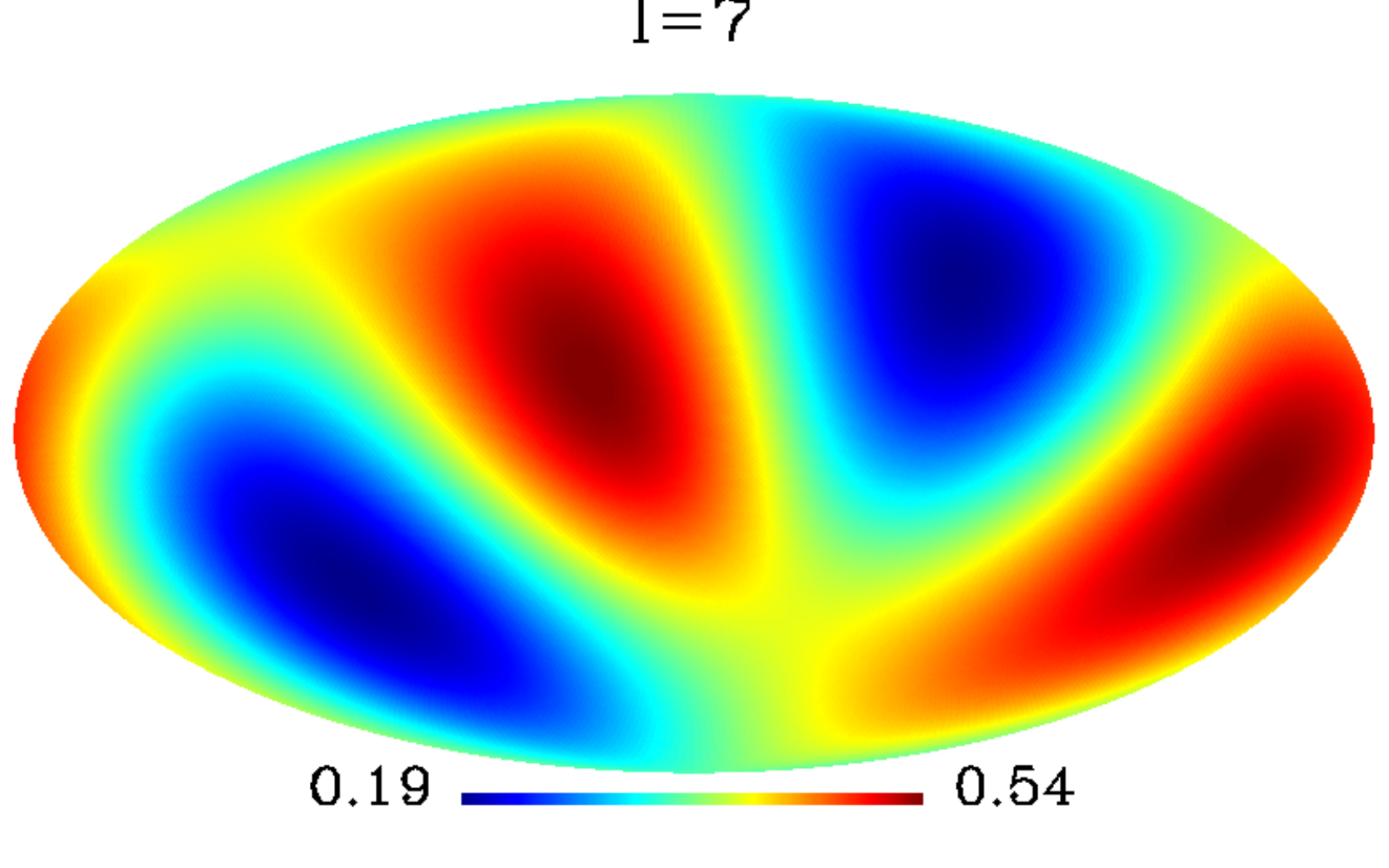}\includegraphics[width=5cm]{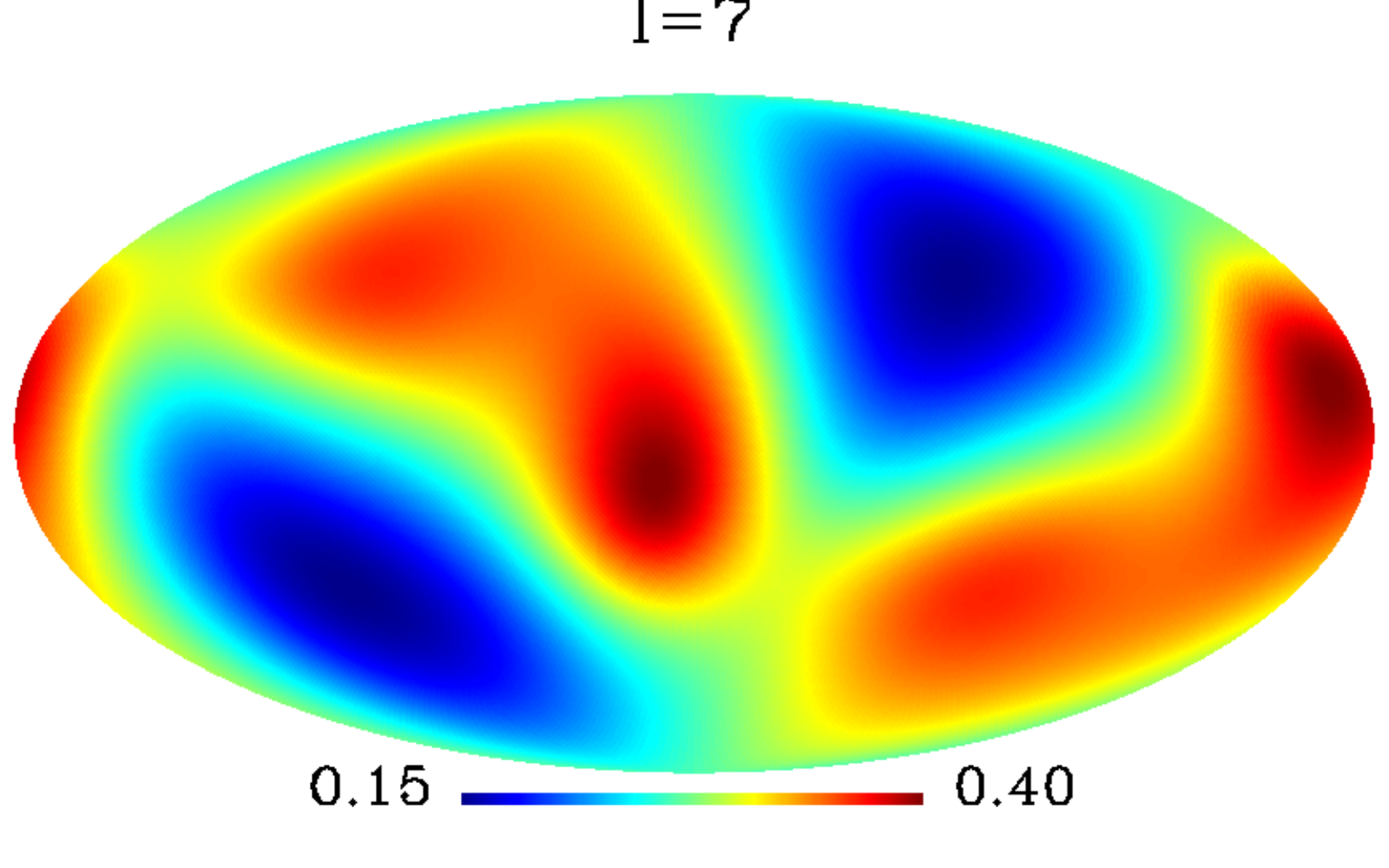} \\
\includegraphics[width=5cm]{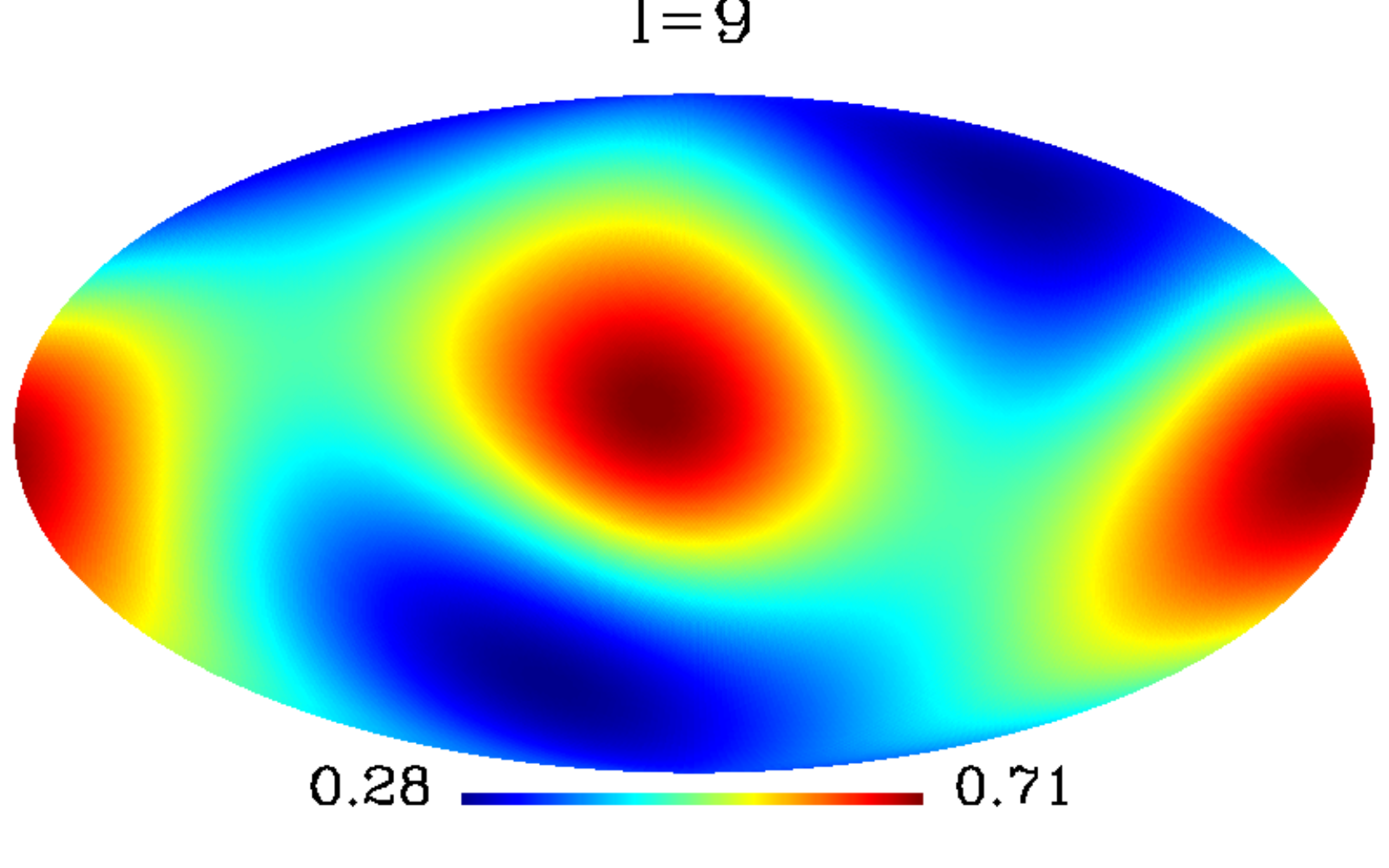}\includegraphics[width=5cm]{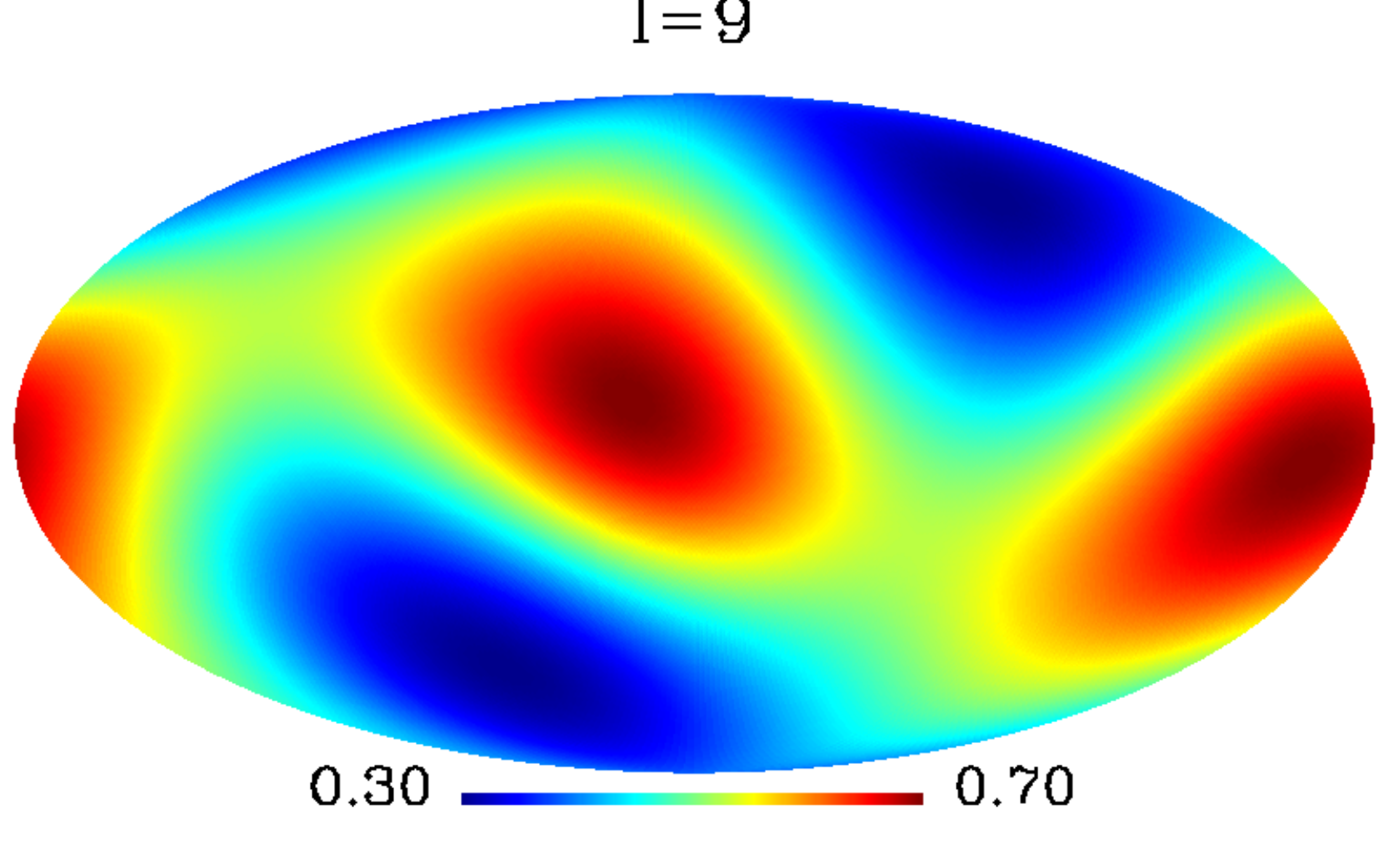}\includegraphics[width=5cm]{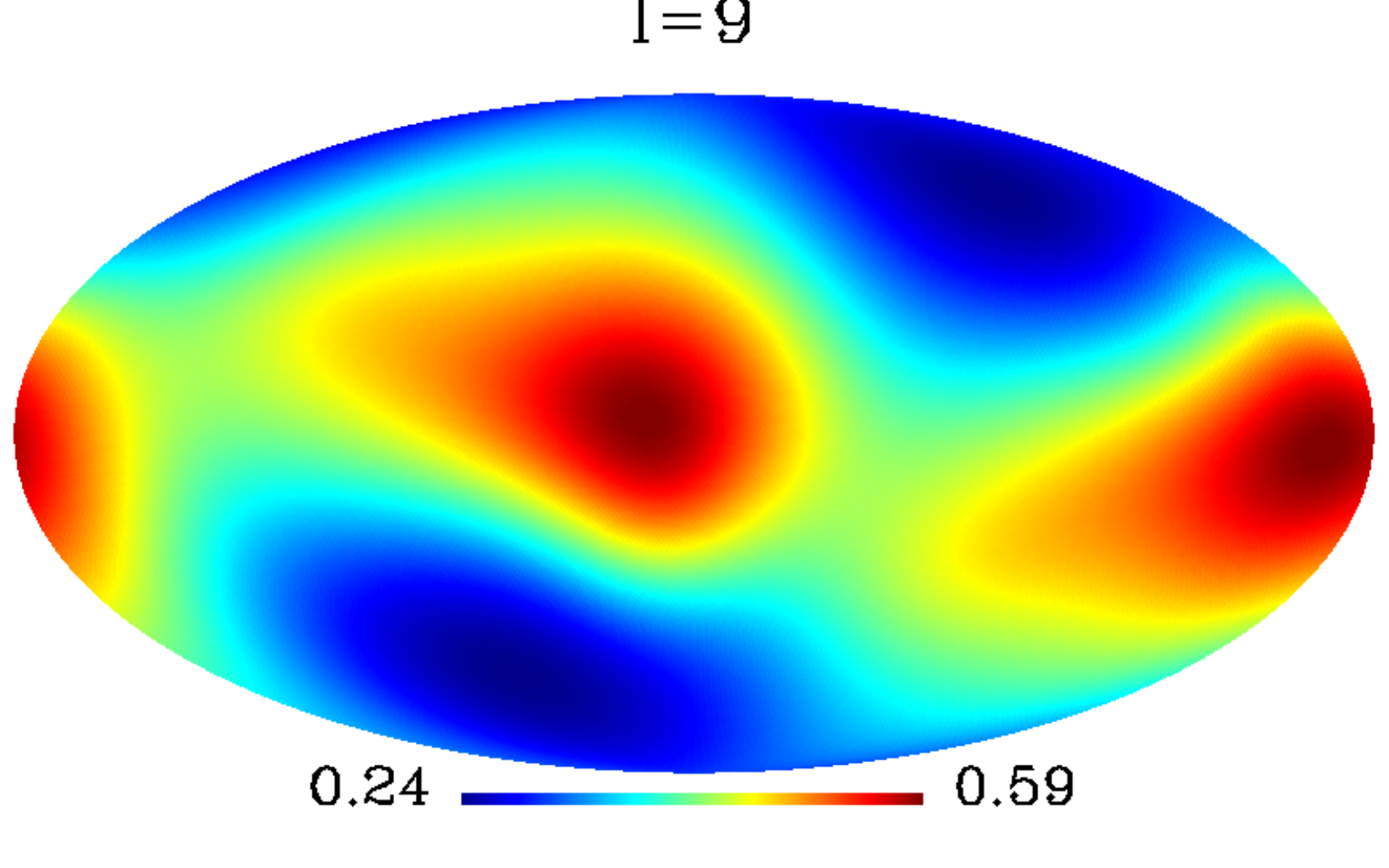}
\end{center}
\caption{Three directional statistics $g_4(l,\hat{\bf q})$ (left),
$g_5(l,\hat{\bf q})$ (middle), and $g_6(l,\hat{\bf q})$ (right) as
functions of $\hat{\bf q}\equiv (\theta,\phi)$. Note that,
these results are based on the Planck SMICA data.}\label{fig2}
\end{figure*}

We apply these three statistics to the Planck SMICA, NILC, and
SEVEM data and find similar results. In Fig.\ref{fig2}, we
present the results of SMICA data, which show that $g_i
(l,\hat{\bf q})< 1$ ($i=4,5,6$) are held for any direction
$\hat{\bf q}$. So we conclude that the odd-parity preference exists even if the
estimator $\tilde{D}_{l}$ is considered.

\section{Correlation with the preferred axes of other CMB observations \label{section3}}

Now, let us investigate how the parity statistics depend on the
direction $\hat{\bf q}$. From Fig.\ref{fig1}, we find that
all the statistics $g_i(l,\hat{\bf q})$ of $i=1,2,3$ have a
quite similar morphology for $l>3$, although the weights of the
multipoles in their definitions are different. The similar results have
also been found for the statistics $g_i(l,\hat{\bf q})$ of
$i=4,5,6$ in Fig.\ref{fig2}. In particular, from these two figures, we find that the panels of $l>3$ have the similar preferred directions ${\bf \hat{q}}$, where the corresponding parity statistic is minimized. These have also been clearly shown in Tables II and III. Meanwhile, the same results are also confirmed in NILC and SEVEM maps \footnote{Similar to Ref. \cite{naselskyzhao}, in both figures, we find that the
morphologies of $l=3$ maps are different from others, which may
connect with the detailed origin of the anomalous CMB quadrupole and
octopole and needs further investigation.}.

\begin{figure*}[t]
\begin{center}
\includegraphics[width=7cm]{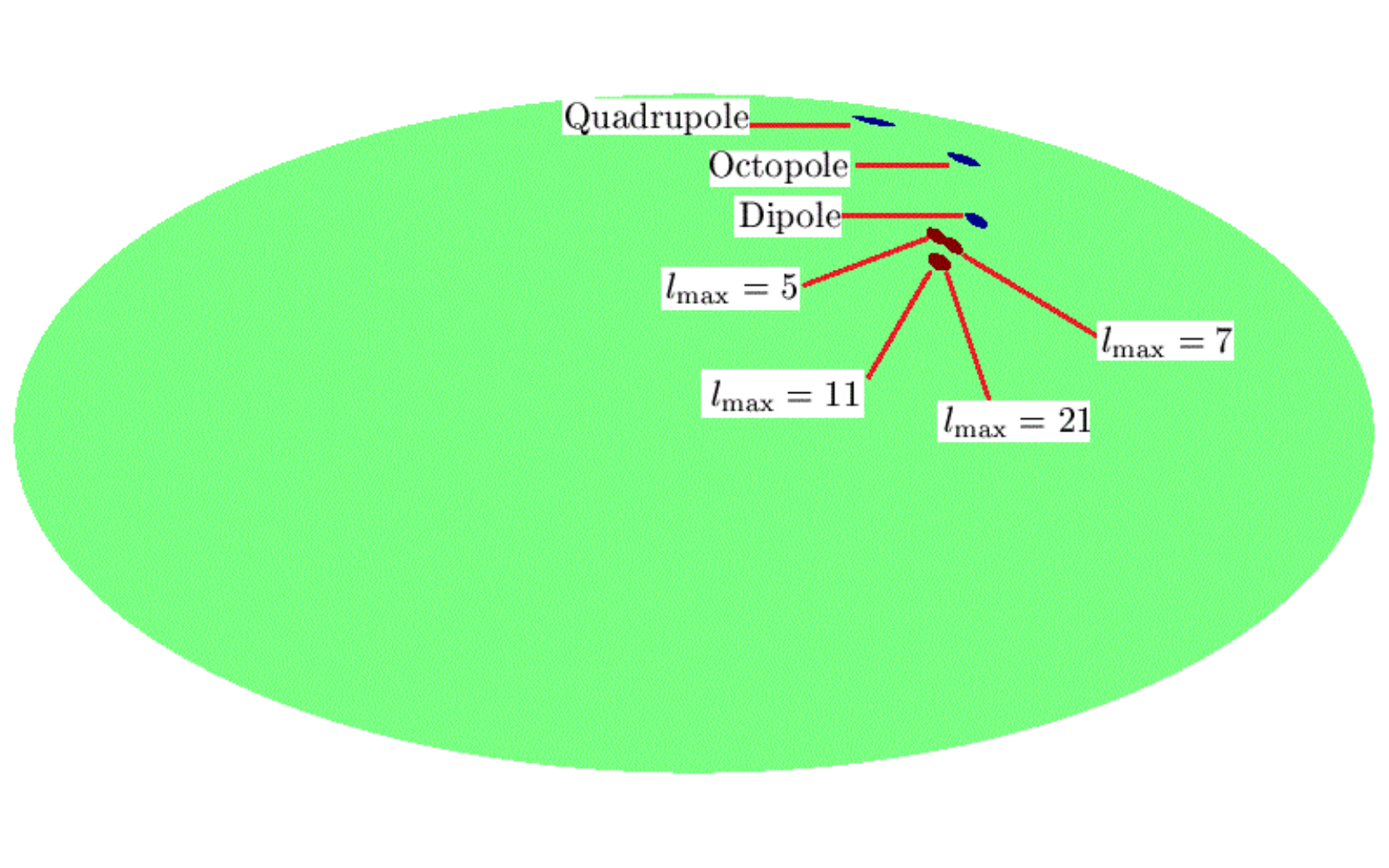}\includegraphics[width=7cm]{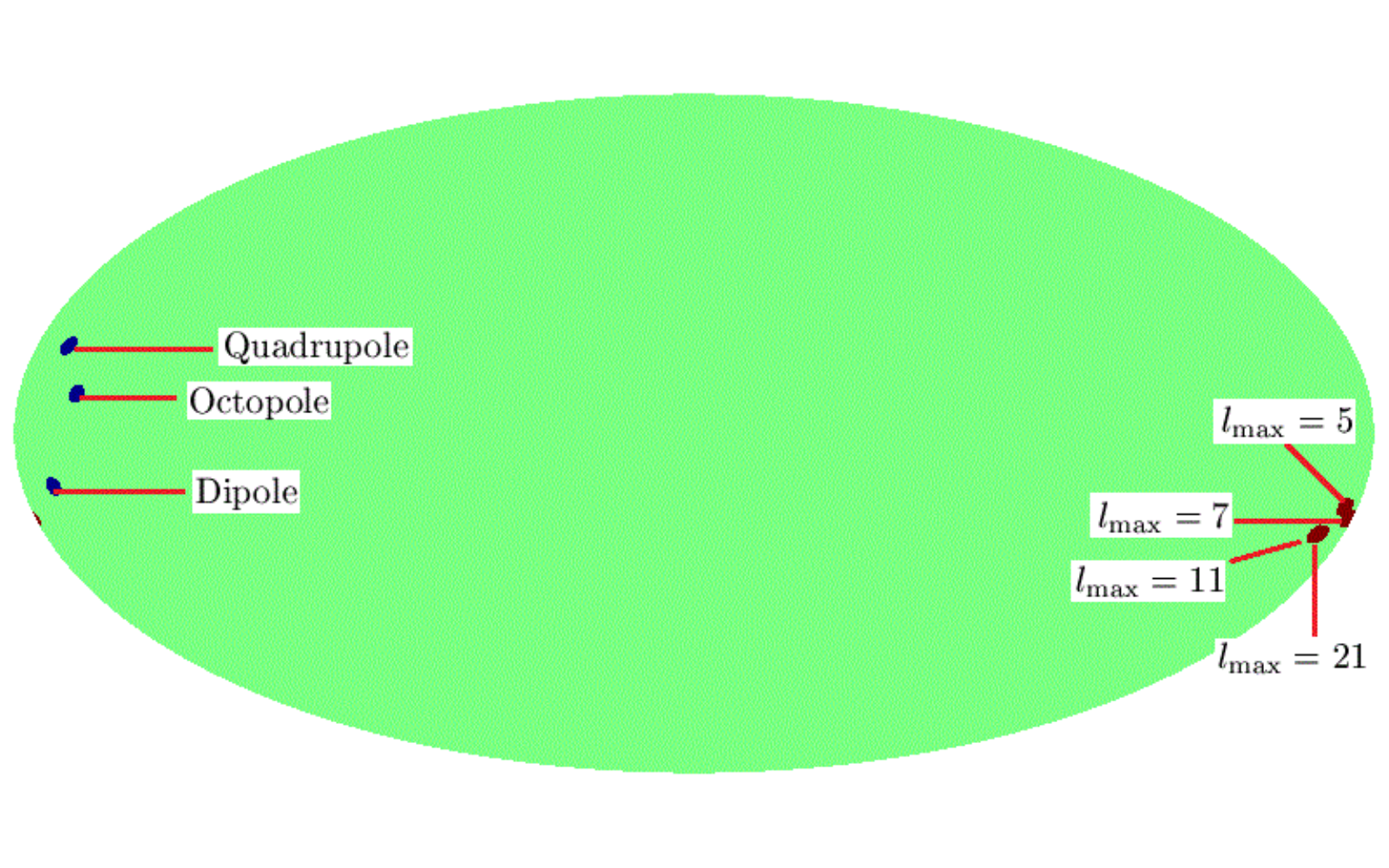}
\end{center}
\caption{The preferred directions of the SMICA-based statistics
$g_1(l,\hat{\bf q})$ in the Galactic coordinate
system (left) and in the ecliptic coordinate system (right). In both
panels, we have compared them with the CMB kinematic dipole
direction and the preferred directions of the CMB quadrupole and
octopole.}\label{fig3}
\end{figure*}

In Fig.\ref{fig3}, we compare the preferred directions $\hat{\bf
q}$ of parity asymmetry with the CMB kinematic dipole and find
they are very close to each other. Especially, all these directions are close to the ecliptic
plane. To quantify it, we define the quantity $\alpha$, which is
the angle between $\hat{\bf q}$ and the CMB kinematic dipole
direction at $(\theta=42^{\circ},\phi=264^{\circ})$ \cite{dipole}.
We list the values of $|\cos\alpha|$ in Tables
\ref{tab1}-\ref{tab4} and find that all of them are very close
to $1$. Note that, throughout this paper we do not differentiate
the direction $\hat{\bf q}$ and the opposite one $-\hat{\bf q}$.

In both WMAP and Planck data, the alignment between the
orientation of the CMB quadrupole and octopole has been reported. In
the Planck SMICA data, the preferred direction of the quadrupole is at
$(\theta=13.4^{\circ}, \phi=238.5^{\circ})$, and that of octopole
is $(\theta=25.7^{\circ}, \phi=239.0^{\circ})$ \cite{quadrupole}.
The angle between them is $12.3^{\circ}$, and the significance of
alignment is $96.8\%$. Such a level may not necessarily correspond to
a statistical significance. However, when combined with the axes directions of other
cosmological observations, such as the anisotropy of cosmic acceleration, the bulk velocity flow axis,
and the quasar optical polarization alignment axis, the statistical evidence of the relative coincidence
could increase dramatically (see, for instance, Ref. \cite{anto}).

In this paper, we shall investigate whether or not
the alignment of quadrupole and octopole connects with the parity
asymmetry. Following Ref. \cite{anto}, it is
straightforward to evaluate the mean value of the inner product
between all the pairs of unit vectors corresponding to the
following four directions: the preferred directions of quadrupole,
octopole, parity asymmetry and the direction of the CMB kinematic
dipole. So, we define the quantity,
 \be\label{ctheta}
 \langle |\cos\theta_{ij}|\rangle = \sum_{i,j=1,~j\neq i}^{N} \frac{|\hat{r}_i\cdot \hat{r}_j|}{N(N-1)},
 \ee
where $N$ is the number of the directions, which will be
investigated. First, we shall study the case in the absence of the
parity asymmetry. The alignment between these three axes was also
reported in WMAP data \cite{schwarz}. In this case, we have $N=3$
and $\langle |\cos\theta_{ij}|\rangle=0.9242$ for the real data.
{To evaluate the significance of the alignment, we pixelize the two-dimensional sphere in the HEALPix format with the resolution parameter $N_{\rm side}=256$, which corresponds to the total pixel number $N_{\rm pix}=12\times N_{\rm side}^2$. Then, we randomly generate $10^5$ realizations. For each realization, the three directions are randomly and independently picked on the sky using a uniform distribution between 0 and $N_{\rm pix}-1$,} and the corresponding value $\langle |\cos\theta_{ij}|\rangle$ is calculated directly. Considering all the random samples, we obtain that $\langle |\cos\theta_{ij}| \rangle = 0.500 \pm 0.167$. {To quantify the significant level of the deviation from the random distribution, we define the
$\Delta_c/\sigma_c$, where $\Delta_c$ is the difference between the observed value of $\langle |\cos\theta_{ij}|\rangle$ and the mean value of the simulations, and $\sigma_c$ is the corresponding standard deviation of the simulations. Considering the observed result $\langle |\cos\theta_{ij}|\rangle=0.9242$, and the simulated value $\sigma_c=0.167$, we obtain that $\Delta_c/\sigma_c=2.54$, which indicates that the alignment of these three directions is around $2.5\sigma$ confidence level.}

Now, let us take into account the preferred direction of the CMB
parity asymmetry. For the $10^5$ random realizations of four
random points on the sphere, by a similar analysis, we get
 \be\label{theta4}
 \langle |\cos\theta_{ij}| \rangle = 0.500 \pm 0.118.
 \ee
As anticipated, compared with the case of $N=3$, the mean value
stays the same, while the standard deviation $\sigma_c$ significantly decreases. Now, we
calculate the real value of the quantity $\langle
|\cos\theta_{ij}| \rangle$. For the preferred direction of parity
asymmetry, we consider the results of all the six statistics
$g_i(l,\hat{\bf q})$ defined in this paper, and list the
corresponding values in Tables \ref{tab1}-\ref{tab4}. For every
case with $l>3$, we find that $\langle |\cos\theta_{ij}| \rangle$
is close to $0.9$, and the corresponding significance of the
alignment between these directions increases to $\Delta_c/\sigma_c \gtrsim 3$.

We therefore conclude that the preferred direction of the CMB
parity asymmetry is not only very close to the CMB kinematic
dipole, but also close to the preferred axes of the CMB quadrupole
and octopole, which is nearly independent of the choice of the
parity statistic. Their coexistence in a relatively small angular
region is a very unlikely event, which implies that these
anomalies have a common origin: an undiscovered physical effect or
a common basic systematic error that has so far escaped attention.

\begin{table*}
\caption{The preferred direction $\hat{\bf q}=(\theta,\phi)$,
where the parity parameter $g_i(l,\hat{\bf q})$ based on Planck
SMICA data is minimized, is compared with the other CMB preferred
axes. In each box, the upper one is the result for the statistic with $i=1$,
the middle one is that for $i=2$, and the lower one is that for $i=3$. In
this table, $\alpha$ is the angle between $\hat{\bf q}$ and the
CMB kinematic dipole, $\langle|\cos\theta_{ij}|\rangle$ is the
quantity defined in Eq. (\ref{ctheta}), and the $\Delta_c/\sigma_c$ value denotes
the number of $\sigma_c$ the observed
$\langle|\cos\theta_{ij}|\rangle$ deviate from the simulations.}
\begin{center}
\label{tab1}
\begin{tabular}{ |c||c |c |c |c |c| }
    \hline
     & ~~~~~~$\theta[^{\circ}]$~~~~~~ & ~~~~~~$\phi[^{\circ}]$~~~~~~ & ~~~~~~$|\cos\alpha|$~~~~~~ & ~~~~$\langle|\cos\theta_{ij}|\rangle$~~~~ & ~~~~~~~$\Delta_c/\sigma_c$~~~~~~~  \\
   \hline
   \hline
   \multirow{3}{*}{$l_{\max}=3$} &    90.00  &  23.20    & 0.3265 &    0.6066        & 0.90 \\
   &    90.00  &  23.20    & 0.3265 &    0.6066        & 0.90 \\
   &    90.00  &  23.20    & 0.3265 &    0.6066        & 0.90 \\
   \hline
   \multirow{3}{*}{$l_{\max}=5$} &   45.80  &   281.07   & 0.9767 &     0.9015    & 3.40 \\
   &   45.80  &   281.07   & 0.9767 &     0.9015    & 3.40 \\
   &   45.80  &   281.07   & 0.9767 &     0.9015    & 3.40 \\
   \hline
   \multirow{3}{*}{$l_{\max}=7$} &   48.19  &   277.73   & 0.9799 &     0.8979    & 3.37 \\
   &   47.39  &   279.29   & 0.9782 &     0.8987    & 3.38 \\
   &   52.83  &   267.89   & 0.9710 &     0.8915    & 3.32 \\
   \hline
   \multirow{3}{*}{$l_{\max}=11$} &   52.08  &   284.06   & 0.9525 &     0.8744    & 3.17 \\
   &   49.77  &   280.54   & 0.9697 &     0.8886    & 3.29 \\
   &   53.58  &   226.41   & 0.8679 &     0.8793    & 3.21 \\
   \hline
   \multirow{3}{*}{$l_{\max}=21$} &   52.08  &   285.47   & 0.9479 &     0.8721    & 3.15 \\
   &   50.55  &   284.06   & 0.9575 &     0.8804    & 3.22 \\
   &   21.32  &   131.90   & 0.5292 &     0.8295    & 2.79 \\
   \hline
\end{tabular}
\end{center}
\end{table*}

\begin{table*}
\caption{The preferred direction $\hat{\bf q}=(\theta,\phi)$,
where the parity parameter $g_i(l,\hat{\bf q})$ based on Planck
SMICA data is minimized, is compared with the other CMB preferred
axes. In each box, the upper one is the result for the statistic with $i=4$,
the middle one is that for $i=5$, and the lower one is that for $i=6$. In
this table, $\alpha$ is the angle between $\hat{\bf q}$ and the
CMB kinematic dipole, $\langle|\cos\theta_{ij}|\rangle$ is the
quantity defined in Eq. (\ref{ctheta}), and the $\Delta_c/\sigma_c$ value denotes
the number of $\sigma_c$ the observed
$\langle|\cos\theta_{ij}|\rangle$ deviate from the simulations.}
\begin{center}
\label{tab2}
\begin{tabular}{|c||c |c |c |c |c| }
    \hline
     & ~~~~~~$\theta[^{\circ}]$~~~~~~ & ~~~~~~$\phi[^{\circ}]$~~~~~~ & ~~~~~~$|\cos\alpha|$~~~~~~ & ~~~~$\langle|\cos\theta_{ij}|\rangle$~~~~ & ~~~~~~~$\Delta_c/\sigma_c$~~~~~~~  \\
   \hline
   \hline
   \multirow{3}{*}{$l_{\max}=3$} &    88.81  &  23.20    & 0.3109 &    0.5975        & 0.83 \\
   &    88.81  &  23.20    & 0.3109 &    0.5975        & 0.83 \\
   &    88.81  &  23.20    & 0.3109 &    0.5975        & 0.83 \\
   \hline
   \multirow{3}{*}{$l_{\max}=5$} &   47.39  &   307.86   & 0.8582 &     0.8458    & 2.93 \\
   &   47.39  &   310.71   & 0.8408 &     0.8390    & 2.87 \\
   &   46.59  &   309.92   & 0.8488 &     0.8442    & 2.92 \\
   \hline
   \multirow{3}{*}{$l_{\max}=7$} &   62.72  &   280.55   & 0.9107 &     0.8306    & 2.80 \\
   &   56.49  &   281.25   & 0.9431 &     0.8599    & 3.05 \\
   &   55.77  &   281.95   & 0.9443 &     0.8621    & 3.07 \\
   \hline
   \multirow{3}{*}{$l_{\max}=9$} &   32.60  &   236.25   & 0.9451 &     0.9424    & 3.75 \\
   &   36.43  &   248.88   & 0.9815 &     0.9418    & 3.74 \\
   &   34.89  &   246.06   & 0.9737 &     0.9435    & 3.76 \\
   \hline

\end{tabular}
\end{center}
\end{table*}

\begin{table*}
\caption{The values of $|\cos\alpha|$ and $\Delta_c/\sigma_c$ for the
statistics $g_i(l,\hat{\bf q})$ based on Planck NILC and SEVEM
data. Similar to Table \ref{tab1}, in each box, the upper one is the result
for the statistic with $i=1$, the middle one is that for $i=2$, and the
lower one is that for $i=3$.}
\begin{center}
\label{tab3}
\begin{tabular}{ |c||c |c| |c |c|  }
    \hline
     & ~~~$|\cos\alpha|$ for NILC~~~ & ~~~$\Delta_c/\sigma_c$ for NILC~~~ & ~~~$|\cos\alpha|$ for SEVEM~~~ & ~~~$\Delta_c/\sigma_c$ for SEVEM~~~   \\
   \hline
   \hline
   \multirow{3}{*}{$l_{\max}=3$} &    0.3259  &  0.88    & 0.2656 &    0.66      \\
   &    0.3259  &  0.88    & 0.2656 &    0.66      \\
   &    0.3259  &  0.88    & 0.2656 &    0.66      \\
   \hline
   \multirow{3}{*}{$l_{\max}=5$} &   0.9758  &   3.38   & 0.9802 &     3.42   \\
   &   0.9748  &   3.36   & 0.9802 &     3.42   \\
   &   0.9748  &   3.36   & 0.9802 &     3.42    \\
   \hline
   \multirow{3}{*}{$l_{\max}=7$} &   0.9822  &   3.37   & 0.9840 &     3.41    \\
   &   0.9769  &   3.36   & 0.9812 &     3.39    \\
   &   0.9812  &   3.33   & 0.9861 &     3.37    \\
   \hline
   \multirow{3}{*}{$l_{\max}=11$} &   0.8600  &   3.18   & 0.8600 &     3.18    \\
   &   0.9697  &   3.29   & 0.9766 &     3.34   \\
   &   0.8520  &   3.15   & 0.8600 &     3.18     \\
   \hline
   \multirow{3}{*}{$l_{\max}=21$} &   0.8932  &   3.25   & 0.9529 &     3.19     \\
   &   0.9575  &   3.22   & 0.9575 &     3.22    \\
   &   0.8522  &   3.13   & 0.5295 &     2.79   \\
   \hline
\end{tabular}
\end{center}
\end{table*}

\begin{table*}
\caption{The values of $|\cos\alpha|$ and $\Delta_c/\sigma_c$ for the statistics $g_i(l,\hat{\bf q})$ based on Planck NILC and SEVEM data. Similar to Table \ref{tab2},
in each box the upper one is the result for the statistic with $i=4$, the middle one is that for $i=5$ and the lower one that is for $i=6$.}
\begin{center}
\label{tab4}
\begin{tabular}{ |c||c |c| c |c|  }
    \hline
     & ~~~$|\cos\alpha|$ for NILC~~~ & ~~~$\Delta_c/\sigma_c$ for NILC~~~ & ~~~$|\cos\alpha|$ for SEVEM~~~ & ~~~$\Delta_c/\sigma_c$ for SEVEM~~~   \\
   \hline
   \hline
   \multirow{3}{*}{$l_{\max}=3$} &    0.3094  &  0.78    & 0.2650 &    0.64      \\
   &    0.3094  &  0.78    & 0.2650 &    0.64      \\
   &    0.3094  &  0.78    & 0.2650 &    0.64      \\
   \hline
   \multirow{3}{*}{$l_{\max}=5$} &   0.8705  &   3.06   & 0.9040 &     3.16   \\
   &   0.8705  &   3.06   & 0.8963 &     3.13   \\
   &   0.8617  &   3.03   & 0.8963 &     3.13    \\
   \hline
   \multirow{3}{*}{$l_{\max}=7$} &   0.9838  &   3.37   & 0.9585 &     3.14    \\
   &   0.9852  &   3.43   & 0.9710 &     3.26    \\
   &   0.9782  &   3.38   & 0.9693 &     3.28    \\
   \hline
   \multirow{3}{*}{$l_{\max}=9$} &   0.9097  &   3.73   & 0.9351 &     3.74    \\
   &   0.9450  &   3.77   & 0.9713 &     3.77   \\
   &   0.9290  &   3.75   & 0.9529 &     3.77   \\
   \hline
\end{tabular}
\end{center}
\end{table*}

\section{Conclusions \label{sction4}}

Parity violation is one of the anomalies of the CMB temperature anisotropy map in the large scales, which may indicate the
nontrivial topology of the Universe, the physics of the early
inflation, the foreground residuals, or some unsolved systematic errors. In this paper,
we have extended our previous work on the directional properties
of the CMB parity violation and applied to the newly released Planck
data. We have defined two kinds of rotationally variant power
spectra $D_l(\hat{\bf q})$ and $\tilde{D}_l(\hat{\bf q})$, where
the special direction had been selected in the definitions.
Based on these estimators, we considered six different parity
statistics, and demonstrated the direction preferences of the CMB
parity asymmetry in the low multipoles.

Similar to the statistics based on the rotationally invariant
estimator $C_l$, we found all these statistics show the odd-parity
preference of the CMB data. At the same time, we found that the preferred directions
$\hat{\bf q}$ of all the statistics, where the statistics are
minimized, are very close to each other as long as $l>3$. In
particular, these preferred directions are close to both the
direction of the CMB kinematic dipole and the preferred directions
of the CMB quadrupole and octopole. The confident level of the deviation
from the random distribution between them is $\Delta_c/\sigma_c \gtrsim 3$, which
implies that the parity violation and the anomalies of the CMB
quadrupole and octopole might have the common dipole-related origin.
This origin might be in physics, in contamination or in systematics. In any case, the
future polarization data, $TE$, $EE$, and $BB$ would be helpful to
resolve the puzzles and the coincidence between them.

In the end, it is interesting to mention that several preferred
axes have also been reported in other cosmological observations:
velocity flows \cite{velocity}, quasar alignment \cite{quasar},
anisotropies of the cosmic acceleration \cite{acceleration,cai0},
{\tc{the dipole in the handedness of spiral galaxies \cite{longo},}}
and the dipole effect of the fine structure constant \cite{fine} (see
Ref. \cite{review} as a review). Although there is still some debate
\cite{kalus,cai,zhao2}, it has been claimed that these preferred
directions are also aligned with the CMB kinematic dipole and the
preferred direction of CMB quadrupole and octopole \cite{review}.
If all the directional preferences would be confirmed in the
future, it would imply that the underlying physical or
systematic reasons of all the cosmological anomalies should be
connected.


{\bf Acknowledgements:}  We acknowledge the use of the Planck
Legacy Archive (PLA). Our data analysis made the use of HELAPix
\cite{healpix} and GLESP \cite{glesp}. This work is supported by Project 973 under Grant No.2012CB821804; NSFC under Grants
No.11173021, No.11075141, No.11322324; and a project of KIP
of CAS.

\end{document}